\documentclass[twocolumn,numberedappendix,twocolappendix,appendixfloats]{openjournal_dhayaa}

\usepackage{amsmath}
\usepackage{fontawesome5}
\usepackage{color}
\usepackage{xcolor}
\usepackage{ulem}

\usepackage{natbib}
\setcitestyle{aysep={}}

\usepackage[colorlinks=true
  ,urlcolor=blue
  ,anchorcolor=blue
  ,citecolor=blue
  ,filecolor=blue
  ,linkcolor=blue
  ,menucolor=blue
  ,linktocpage=true
  ,pdfproducer=medialab
  ,pdfa=true
]{hyperref}

\usepackage{graphicx}	
\usepackage{amsmath}	

\usepackage{fontawesome5}
\usepackage{color}
\usepackage{xcolor}
\usepackage{xspace}

\usepackage{enumitem}
\setlist{nosep}

\newcommand{\eg}{{\sl e.g.}, }     
\newcommand{\ie}{{\sl i.e.}, }

\newcommand{\Gaia}{\textit{Gaia}\xspace}
\newcommand{\decade}{DECADE\xspace}

\definecolor{orcidlogocol}{HTML}{A6CE39}
\definecolor{purple}{RGB}{128, 0, 128}
\definecolor{kelly}{RGB}{76, 187, 23}

\newcommand{\OrcidID}[1]{ \href[urlcolor = red]{https://orcid.org/#1}{\textcolor{lightgray}{\faOrcid}}}
\newcommand{\OrcidIDName}[2]{\href{https://orcid.org/#1}{#2}}

\defcitealias{y3-shapecatalog}{\textsc{GS21}}
\defcitealias{paper1}{\textsc{Paper I}}
\defcitealias{paper2}{\textsc{Paper II}}
\defcitealias{paper3}{\textsc{Paper III}}
\defcitealias{paper4}{\textsc{Paper IV}}

\newcommand*{\vcenteredhbox}[1]{\begingroup
\setbox0=\hbox{#1}\parbox{\wd0}{\box0}\endgroup}

\begin{document}

{\hfill FERMILAB-PUB-25-0063-LDRD-PPD}

\title{The DECADE cosmic shear project I: A new weak lensing shape catalog of 107 million galaxies} 
\shortauthors{Anbajagane \& Chang et. al}
\shorttitle{The DECADE cosmic shear project I: shape catalog}

\author{\OrcidIDName{0000-0003-3312-909X}{D.~Anbajagane} (\vcenteredhbox{\includegraphics[height=1.2\fontcharht\font`\B]{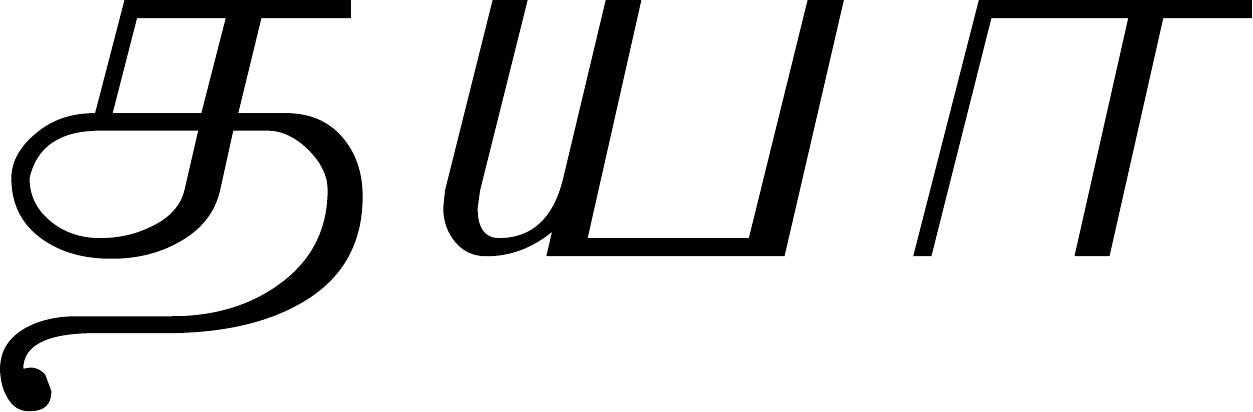}})$^\star$}
\affiliation{Department of Astronomy and Astrophysics, University of Chicago, Chicago, IL 60637, USA}
\affiliation{Kavli Institute for Cosmological Physics, University of Chicago, Chicago, IL 60637, USA}
\email{$^{\star}$dhayaa@uchicago.edu, chihway@kicp.uchicago.edu}

\author{\OrcidIDName{0000-0002-7887-0896}{C.~Chang}$^\star$}
\affiliation{Department of Astronomy and Astrophysics, University of Chicago, Chicago, IL 60637, USA}
\affiliation{Kavli Institute for Cosmological Physics, University of Chicago, Chicago, IL 60637, USA}

\author{\OrcidIDName{0000-0002-7523-582X}{Z.~Zhang}}
\affiliation{Department of Astronomy and Astrophysics, University of Chicago, Chicago, IL 60637, USA}
\affiliation{Department of Physics, Stanford University, 382 Via Pueblo Mall, Stanford, CA 94305, USA}
\affiliation{SLAC National Accelerator Laboratory, Menlo Park, CA 94025, USA}

\author{\OrcidIDName{0000-0003-0478-0473}{C.~Y.~Tan}}
\affiliation{Department of Physics, University of Chicago, Chicago, IL 60637, USA}
\affiliation{Kavli Institute for Cosmological Physics, University of Chicago, Chicago, IL 60637, USA}

\author{\OrcidIDName{0000-0002-6904-359X}{M.~Adamow}}
\affiliation{Center for Astrophysical Surveys, National Center for Supercomputing Applications, 1205 West Clark St., Urbana, IL 61801, USA}
\affiliation{Department of Astronomy, University of Illinois at Urbana-Champaign, 1002 W. Green Street, Urbana, IL 61801, USA}

\author{\OrcidIDName{0000-0002-6002-4288}{L.~F.~Secco}}
\affiliation{Kavli Institute for Cosmological Physics, University of Chicago, Chicago, IL 60637, USA}

\author{\OrcidIDName{0000-0001-7774-2246}{M.~R.~Becker}}
\affiliation{Argonne National Laboratory, 9700 South Cass Avenue, Lemont, IL 60439, USA}

\author{\OrcidIDName{0000-0001-6957-1627}{P.~S.~Ferguson}}
\affiliation{DIRAC Institute, Department of Astronomy, University of Washington, 3910 15th Ave NE, Seattle, WA, 98195, USA}

\author{\OrcidIDName{0000-0001-8251-933X}{A.~Drlica-Wagner}}
\affiliation{Fermi National Accelerator Laboratory, P. O. Box 500, Batavia, IL 60510, USA}
\affiliation{Department of Astronomy and Astrophysics, University of Chicago, Chicago, IL 60637, USA}
\affiliation{Kavli Institute for Cosmological Physics, University of Chicago, Chicago, IL 60637, USA}

\author{\OrcidIDName{0000-0002-4588-6517}{R.~A.~Gruendl}}
\affiliation{Center for Astrophysical Surveys, National Center for Supercomputing Applications, 1205 West Clark St., Urbana, IL 61801, USA}
\affiliation{Department of Astronomy, University of Illinois at Urbana-Champaign, 1002 W. Green Street, Urbana, IL 61801, USA}

\author{\OrcidIDName{0000-0003-4394-7491}{K.~Herron}}
\affiliation{Department of Physics and Astronomy, Dartmouth College, Hanover, NH 03755, USA}

\author{\OrcidIDName{0009-0002-4207-0210}{A.~Tong}}
\affiliation{Department of Physics and Astronomy, University of Pennsylvania, Philadelphia, PA 19104, USA}

\author{\OrcidIDName{0000-0002-5622-5212}{M.~A.~Troxel}}
\affiliation{Department of Physics, Duke University Durham, NC 27708, USA}

\author{\OrcidIDName{0000-0003-3054-7907}{D.~Sanchez-Cid}}
\affiliation{Centro de Investigaciones Energéticas, Medioambientales y Tecnológicas (CIEMAT), Madrid, Spain}
\affiliation{Physik-Institut, University of Zurich, Winterthurerstrasse 190, CH-8057 Zurich, Switzerland}

\author{\OrcidIDName{0000-0002-1831-1953}{I.~Sevilla-Noarbe}}
\affiliation{Centro de Investigaciones Energ\'eticas, Medioambientales y Tecnol\'ogicas (CIEMAT), Madrid, Spain}

\author{\OrcidIDName{0009-0005-1143-495X}{N.~Chicoine}}
\affiliation{Department of Astronomy and Astrophysics, University of Chicago, Chicago, IL 60637, USA}
\affiliation{Department of Physics and Astronomy, University of Pittsburgh, 3941 O’Hara Street, Pittsburgh, PA 15260}

\author{\OrcidIDName{0000-0002-5279-0230}{R.~Teixeira}}
\affiliation{Department of Astronomy and Astrophysics, University of Chicago, Chicago, IL 60637, USA}
\affiliation{Department of Physics, Duke University Durham, NC 27708, USA}

\author{\OrcidIDName{0000-0001-8505-1269}{A.~Alarcon}}
\affiliation{Institute of Space Sciences (ICE, CSIC),  Campus UAB, Carrer de Can Magrans, s/n,  08193 Barcelona, Spain}

\author{\OrcidIDName{0000-0003-2911-2025}{D.~Suson}}
\affiliation{Department of Chemistry and Physics, Purdue University Northwest 2200, 169th Ave, Hammond, IN 46323}

\author{\OrcidIDName{0000-0002-3173-2592}{A.~N.~Alsina}}
\affiliation{Instituto de Física Gleb Wataghin, Universidade Estadual de Campinas, 13083-859, Campinas, SP, Brazil}

\author{\OrcidIDName{0000-0002-6445-0559}{A.~Amon}}
\affiliation{Department of Astrophysical Sciences, Princeton University, Peyton Hall, Princeton, NJ 08544, USA}

\author{\OrcidIDName{0000-0003-4383-2969}{C.~R.~Bom}}
\affiliation{Centro Brasileiro de Pesquisas F\'isicas, Rua Dr. Xavier Sigaud 150, 22290-180 Rio de Janeiro, RJ, Brazil}

\author{\OrcidIDName{0000-0002-3690-105X}{J.~A.~Carballo-Bello}}
\affiliation{Instituto de Alta Investigaci\'on, Universidad de Tarapac\'a, Casilla 7D, Arica, Chile}

\author{\OrcidIDName{0000-0003-1697-7062}{W.~Cerny}}
\affiliation{Department of Astronomy, Yale University, New Haven, CT 06520, USA}

\author{\OrcidIDName{0000-0002-5636-233X}{A.~Choi}}
\affiliation{NASA Goddard Space Flight Center, 8800 Greenbelt Rd, Greenbelt, MD 20771, USA}

\author{\OrcidIDName{0000-0003-1680-1884}{Y.~Choi}}
\affiliation{NSF National Optical-Infrared Astronomy Research Laboratory}

\author{\OrcidIDName{0000-0003-4480-0096}{C.~Doux}}
\affiliation{Université Grenoble Alpes, CNRS, LPSC-IN2P3, 38000 Grenoble, France}

\author{\OrcidIDName{0000-0002-1407-4700}{K.~Eckert}}
\affiliation{Department of Physics and Astronomy, University of Pennsylvania, Philadelphia, PA 19104, USA}

\author{\OrcidIDName{0000-0001-6134-8797}{M.~Gatti}}
\affiliation{Kavli Institute for Cosmological Physics, University of Chicago, Chicago, IL 60637, USA}

\author{\OrcidIDName{0000-0003-3270-7644}{D.~Gruen}}
\affiliation{University Observatory, Faculty of Physics, Ludwig-Maximilians-Universität, Scheinerstr. 1, 81679 Munich, Germany}
\affiliation{Excellence Cluster ORIGINS, Boltzmannstr. 2, 85748 Garching, Germany}

\author{\OrcidIDName{0000-0001-5160-4486}{D.~J.~James}}
\affiliation{Applied Materials Inc., 35 Dory Road, Gloucester, MA 01930}
\affiliation{ASTRAVEO LLC, PO Box 1668, Gloucester, MA 01931}

\author{\OrcidIDName{0000-0002-4179-5175}{M.~Jarvis}}
\affiliation{Department of Physics and Astronomy, University of Pennsylvania, Philadelphia, PA 19104, USA}

\author{\OrcidIDName{0000-0003-2511-0946}{N.~Kuropatkin}}
\affiliation{Fermi National Accelerator Laboratory, P. O. Box 500, Batavia, IL 60510, USA}

\author{\OrcidIDName{0000-0002-9144-7726}{C.~E.~Mart\'inez-V\'azquez}}
\affiliation{International Gemini Observatory/NSF NOIRLab, 670 N. A'ohoku Place, Hilo, Hawai'i, 96720, USA}

\author{\OrcidIDName{0000-0002-8093-7471}{P.~Massana}}
\affiliation{NSF's NOIRLab, Casilla 603, La Serena, Chile}

\author{\OrcidIDName{0000-0003-3519-4004}{S.~Mau}}
\affiliation{Department of Physics, Stanford University, 382 Via Pueblo Mall, Stanford, CA 94305, USA}
\affiliation{Kavli Institute for Particle Astrophysics \& Cosmology, P.O.\ Box 2450, Stanford University, Stanford, CA 94305, USA}

\author{\OrcidIDName{0000-0002-4475-3456}{J.~McCullough}}
\affiliation{Department of Astrophysical Sciences, Peyton Hall, Princeton University, Princeton, NJ USA 08544}

\author{\OrcidIDName{0000-0003-0105-9576}{G.~E.~Medina}}
\affiliation{Dunlap Institute for Astronomy \& Astrophysics, University of Toronto, 50 St George Street, Toronto, ON M5S 3H4, Canada}
\affiliation{David A. Dunlap Department of Astronomy \& Astrophysics, University of Toronto, 50 St George Street, Toronto ON M5S 3H4, Canada}

\author{\OrcidIDName{0000-0001-9649-4815}{B.~Mutlu-Pakdil}}
\affiliation{Department of Physics and Astronomy, Dartmouth College, Hanover, NH 03755, USA}

\author{\OrcidIDName{0000-0001-9438-5228}{M. ~ Navabi}}
\affiliation{Department of Physics, University of Surrey, Guildford GU2 7XH, UK}

\author{\OrcidIDName{0000-0002-8282-469X}{N.~E.~D.~Noël}}
\affiliation{Department of Physics, University of Surrey, Guildford GU2 7XH, UK}

\author{\OrcidIDName{0000-0002-6021-8760}{A.~B.~Pace}}
\affiliation{Department of Astronomy, University of Virginia, 530 McCormick Road, Charlottesville, VA 22904, USA}

\author{\OrcidIDName{0000-0002-5933-5150}{J.~Prat}}
\affiliation{Nordita, KTH Royal Institute of Technology and Stockholm University, SE-106 91 Stockholm.}

\author{\OrcidIDName{0000-0002-7354-3802}{M.~Raveri}}
\affiliation{Department of Physics and INFN, University of Genova, Genova, Italy}

\author{\OrcidIDName{0000-0001-5805-5766}{A.~H.~Riley}}
\affiliation{Institute for Computational Cosmology, Department of Physics, Durham University, South Road, Durham DH1 3LE, UK}

\author{\OrcidIDName{0000-0001-9376-3135}{E.~S.~Rykoff}}
\affiliation{SLAC National Accelerator Laboratory, Menlo Park, CA 94025, USA}
\affiliation{Kavli Institute for Particle Astrophysics \& Cosmology, P.O.\ Box 2450, Stanford University, Stanford, CA 94305, USA}

\author{\OrcidIDName{0000-0002-1594-1466}{J.~D.~Sakowska}}
\affiliation{Department of Physics, University of Surrey, Guildford GU2 7XH, UK}

\author{\OrcidIDName{0000-0003-4102-380X}{D.~J.~Sand}}
\affiliation{Steward Observatory, University of Arizona, 933 North Cherry Avenue, Tucson, AZ 85721-0065, USA}

\author{\OrcidIDName{0000-0003-3402-6164}{L.~Santana-Silva}}
\affiliation{Centro Brasileiro de Pesquisas F\'isicas, Rua Dr. Xavier Sigaud 150, 22290-180 Rio de Janeiro, RJ, Brazil}

\author{\OrcidIDName{0000-0002-6389-5409}{T.~Shin}}
\affiliation{Department of Physics, Carnegie Mellon University, Pittsburgh, PA 15213}

\author{\OrcidIDName{0000-0001-6082-8529}{M.~Soares-Santos}}
\affiliation{Physik-Institut, University of Zurich, Winterthurerstrasse 190, CH-8057 Zurich, Switzerland}

\author{\OrcidIDName{0000-0003-1479-3059}{G.~S.~Stringfellow}}
\affiliation{Center for Astrophysics and Space Astronomy, University of Colorado, 389 UCB, Boulder, CO 80309-0389, USA}

\author{\OrcidIDName{0000-0003-4341-6172}{A.~K.~Vivas}}
\affiliation{Cerro Tololo Inter-American Observatory/NSF NOIRLab, Casilla 603, La Serena, Chile}

\author{\OrcidIDName{0000-0003-1585-997X}{M.~Yamamoto}}
\affiliation{Department of Astrophysical Sciences, Princeton University, Peyton Hall, Princeton, NJ 08544, USA}

\begin{abstract}
We present the Dark Energy Camera All Data Everywhere (\decade) weak lensing dataset: a catalog of 107 million galaxies observed by the Dark Energy Camera (DECam) in the northern Galactic cap. This catalog was assembled from public DECam data including survey and standard observing programs. 
These data were consistently processed with the Dark Energy Survey Data Management pipeline as part of the \decade campaign and serve as the basis of the DECam Local Volume Exploration survey (DELVE) Early Data Release 3 (EDR3). We apply the \textsc{Metacalibration} measurement algorithm to generate and calibrate galaxy shapes. 
After cuts, the resulting cosmology-ready galaxy shape catalog covers a region of $5,\!412 \deg^2$ with an effective number density of $4.59\,\, {\rm arcmin}^{-2}$. The coadd images used to derive this data have a median limiting magnitude of $r = 23.6$, $i = 23.2$, and $z  = 22.6$, estimated at ${\rm S/N} = 10$ in a $2\arcsec$ aperture. We present a suite of detailed studies to characterize the catalog, measure any residual systematic biases, and verify that the catalog is suitable for cosmology analyses. In parallel, we build an image simulation pipeline to characterize the remaining multiplicative shear bias in this catalog, which we measure to be $m = (-2.454 \pm 0.124) \times10^{-2}$ for the full sample. Despite the significantly inhomogeneous nature of the dataset, due to it being an amalgamation of various observing programs, we find the resulting catalog has sufficient quality to yield competitive cosmological constraints.
\end{abstract}


\section{Introduction}

Measurements of weak gravitational lensing --- the deflection of light from distant sources by the intervening matter distribution between the source and the observer --- provide important constraints on the growth, evolution, and content of the Universe \citep{Bartelmann2001, Schneider:2005:Lensing}. 
The cosmological lensing effect, which depends on the gravitational potential field, is seeded by the \textit{total} matter distribution of our Universe. Thus, weak lensing is directly sensitive to \textit{all} matter components, including those that do not emit/absorb light and would otherwise be unobservable. This makes lensing a powerful probe of the underlying structure of the Universe \citep[see][for a review of weak gravitational lensing]{Bartelmann2001} and of any processes that impact this structure; including modified gravity \citep[\eg][]{Schmidt:2008:MG_WL}, primordial signatures \citep[\eg][]{Anbajagane2023Inflation, Goldstein:2024:inflation, Primordial1, Primordial2}, as well as a wide variety of astrophysical impacts \citep[\eg][]{Chisari2018BaryonsPk, Schneider2019Baryonification, Arico:2021:Bacco, Grandis:2024:XrayLensing, Bigwood:2024:BaryonsWLkSZ,  Anbajagane:2024:Baryonification}.

The primary observable for weak lensing is the \textit{shapes} (or more precisely, the \textit{ellipticities}) of galaxies. Since the first detection of weak lensing more than two decades ago \citep{Bacon2000,Kaiser2000,Wittman2000}, the cosmology community has invested significant effort in increasing the statistical power of, and reducing the systematic biases in, these measurements. 

At the heart of these advances are increasingly larger and higher-quality datasets, which have consistently grown in sky coverage, depth, and image quality. The community has now evolved from the early weak lensing surveys that have a few million source galaxies\footnote{Throughout this work, we follow common nomenclature used by the community in referring to galaxies used in the weak lensing measurement as ``source galaxies''.}, such as the Canada-France-Hawaii Telescope Lensing Survey \citep[CFHTLenS,][]{Heymans2013} and the Deep Lens Survey \citep[DLS,][]{Jee2013}, to current Stage-III\footnote{The ``Stage-N'' terminology was introduced in \cite{Albrecht2006} to describe the different phases of dark energy experiments.  There are currently 4 stages, where Stage-III refers to the dark energy experiments that started in the 2010s and Stage-IV refers to those that start in the 2020s.} surveys that have tens to a hundred million source galaxies, such as the Kilo-Degree Survey \citep[KiDS,][]{deJong2015}, the Hyper Suprime-Cam Subaru Strategic Program \citep[HSC-SSP,][]{Aihara2018a}, and the Dark Energy Survey \citep[DES,][]{DES2018}. Other datasets, such as the Ultra-violet NearInfrared Optical Northern Survey (UNIONS) are also building source-galaxy catalogs \citep{Guinot:2022:WL_UNIONS}. In the near future, we expect to observe more than a billion source galaxies with the Vera C.\ Rubin Observatory's Legacy Survey of Space and Time \citep[LSST,][]{Abell2009}, as well as with the \textit{Euclid} space telescope \citep{Laureijs:2011:Euclid}. 
Alongside increases in the statistical power of surveys, there have been significant advances in the methodologies used to measure the shapes of a large number of faint, distant galaxies \citep[e.g.,][]{Bridle2010, Kitching2012, Mandelbaum2015}. Currently, shape measurement techniques are reasonably mature and well-understood, but they continue to be improved \citep{Bernstein2014, Sheldon2020, Li2023}. These improvements help extract more information from the data and tackle increasingly subtle and complex systematic effects that become more apparent with improvements in statistical precision, data quality, and survey depth.

While we have mentioned many of the current, state-of-the-art lensing surveys above, this list does not encapsulate all of the high-quality \textit{image data} that is currently available for cosmological analyses. In particular, DES has shown that the Dark Energy Camera \citep[DECam,][]{Flaugher2015b} on the 4-meter Blanco Telescope in Chile is a high-precision system for measuring cosmology from weak lensing. DES used DECam to perform a homogeneous survey covering 5,000\,deg$^2$ of sky in 760 distinct full- and half-nights from 2013 to 2019 \citep{DES2021}. However, the survey only consists of approximately one-sixth of the data that DECam collected during its nearly continuous operation across more than a decade \citep{Vivas:2023}.

Here, we present the Dark Energy Camera All Data Everywhere (\decade) weak lensing dataset, a new shape catalog derived from multi-band DECam imaging in sky area completely outside of the DES footprint. The resulting cosmology-ready sample contains $107$ million galaxies assembled from $5{,}412$\,deg$^2$ of DECam imaging in the northern Galactic cap that does not overlap with DES\footnote{All areas in this paper are calculated using a \textsc{HEALPix} map \citep{Gorski:2005:Healpix} of \texttt{NSIDE}=4096, with a binary mask (the full pixel area is counted as long as there is one galaxy in that pixel).}.
Compared to the shape catalog from the first three years (Y3) of DES data \citep{y3-shapecatalog}, our new galaxy shape catalog covers $\approx\! 1,000$ deg$^2$ more sky area but with less optimal image quality and uniformity. In particular, DES was designed to optimize the homogeneity of the image depth/quality \citep{Neilsen:2019,Diehl:2019}, while the dataset used in this work is assembled from a wide variety of DECam survey programs and smaller community observations with disparate science goals. Nonetheless, the exceptional field of view and image quality of DECam, coupled with the large collecting area of the 4-meter Blanco Telescope, make the \decade shape catalog a powerful dataset for cosmology. In particular, our catalog has three million more galaxies than the one from DES Y3 \citep{y3-shapecatalog} and as a result the two datasets have similar statistical power.

In order to confidently leverage the DECam data for weak lensing analysis, we apply mature and well-vetted pipelines for data reduction and shape measurement to the imaging data. 
We process the community DECam data with the DES Data Management (DESDM) pipeline \citep{Morganson:2018} at the National Center for Supercomputing Applications (NCSA) as part of the \decade data processing campaign. This includes the full complexity of the calibration products and algorithms developed by the DES Collaboration to correct for instrumental effects \citep{Gruen:2015:BF, Bernstein:2017:Detrending, Bernstein:2018:Photometry, Morganson:2018}. Following DES Y3, we apply the well-characterized \textsc{Metacalibration} shear measurement algorithms \citep{Sheldon2017, Huff2017} to measure shapes of the detected sources. Wherever possible, we follow the analysis procedures and validation processes developed for the DES cosmology analysis. As such, our catalog and subsequent shear analysis present a novel cross check on the DES comsic shear cosmology analysis as we apply very similar algorithms and analysis choices to a completely independent catalog of comparable statistical precision.

The \decade shape catalog has some unique features that make it especially valuable given the recent trends seen in weak lensing datasets, where the measurement of the cosmological parameter $\sigma_{8}$ (defined as the normalization of the matter fluctuations on $8h^{-1}$Mpc scales) is systematically low compared to that from the CMB data \citep{Asgari2021, Amon2022, Secco2022, Dalal2023, Wright:2025:KidsCosmo}. Primarily, a majority of our data covers a region of the sky that has no overlap with other Stage-III lensing surveys and thus provides an important, independent statistical sample for measuring $\sigma_8$. 

In addition, our data will be fairly compatible with the DES data (given the similarity in the processing pipeline and other analysis choices), making it easier to compare to, and combine with, the lensing cosmology constraints from DES \citep[following \eg][]{Chang2019, Longley2023, DESKiDS2023}.
Finally, our data covers a region of the sky that has significant overlap with large spectroscopic surveys such the Sloan Digital Sky Survey \citep{York_2000} and the Dark Energy Spectroscopic Instrument \citep{DESI:2016:Part1}, as well as ground-based CMB experiments such as the South Pole Telescope \citep{Carlstrom2011}, the Atacama Cosmology Telescope \citep{ACT:2007, ACT:2016}, and the Simons Observatory \citep{Simons:2019:Experiment}. This overlap allows for important systematic checks on photometric redshifts \citep[\eg][]{ Schneider:2006:WZ,Newman:2008:WZ,Menard2013, Cawthon2022, Giannini2022}, intrinsic alignment models \citep[\eg][]{Samuroff2022} as well as for opportunities in cross-correlation analyses for a variety of different science cases \citep[\eg][]{Vikram2017GalaxyGroupstSZ, Omori2019, Shin:2019:Splashback, Pandey2019GalaxytSZ, Chang2023, Sanchez2022Sheary,  Anbajagane:2024:Shocks, Bocquet:2024:Clusters}.

Throughout this paper, we closely follow the analysis choices and methodology used to assemble and test the DES Year 1 (Y1) and Year 3 (Y3) shape catalogs, as described in \citet{Zuntz2018} and \citet[][hereafter \citetalias{y3-shapecatalog}]{y3-shapecatalog}, respectively. This paper is the first in a series of four papers describing the \decade cosmic shear analysis. This paper (hereafter \textsc{Paper I}) describes the shape measurement method, the final cosmology sample, and the robustness tests and image simulation pipeline from which we quantify the accuracy of the shear measurements. The second paper \citep[][hereafter \citetalias{paper2}]{paper2} describes the tomographic bin selections and calibrated redshift distributions for our cosmology sample (including the synthetic source injection pipeline needed for these tasks) alongside a series of validation tests of the redshift distributions. The third paper (\href{\#cite.paper3}{Anbajagane \& Chang et al. \citeyear{paper3}}, hereafter \citetalias{paper3}) defines the modeling choices of the cosmology analysis --- including scale cuts, blinding strategy, covariance validation etc. --- as well as the robustness of our cosmology analysis against a variety of systematics, including the point spread function (PSF) contamination, survey inhomogeneity etc. Finally, the fourth paper (\href{\#cite.paper4}{Anbajagane \& Chang et al. \citeyear{paper4}}, hereafter \citetalias{paper4}) performs the cosmological inference with our cosmic shear measurements.

This paper is structured as follows. In Section~\ref{sec:data}, we briefly describe the processing of the DECam data and detail the construction of the base galaxy catalog. In Section~\ref{sec:shear}, we introduce the \textsc{Metacalibration} algorithm used for the shear estimation, the different selection cuts we apply, and the characteristics of the resulting galaxy shape catalog. In Section~\ref{sec:sys_tests}, we analyze the catalog to quantify any residual systematic uncertainties in our measurements and validate these measurements through a series of null-tests. In Section~\ref{sec:calibration_bias}, we describe the image simulation pipeline used to quantify the residual shear calibration bias in our catalog. We summarize in Section~\ref{sec:summary}.

\section{Data Processing}
\label{sec:data}

\subsection{DELVE Early Data Release 3}
\label{sec:desdm}

The data used here were assembled from public DECam data available by December 2022. The largest contributing programs are the DECam Legacy Survey \citep[DECaLS,][]{Dey:2019}, DECam Local Volume Exploration survey \citep[DELVE,][]{Drlica-Wagner:2021, Drlica-Wagner:2022}, and the DECam eROSITA Survey (DeROSITAS, PI: Zenteno). When combined with other community observations, the public DECam data cover more than 21,000 $\deg^2$  of the high Galactic latitude ($|b| > 10\degr$) sky \citep{Drlica-Wagner:2022}. The data products described here were produced for the DELVE Early Data Release 3 (DELVE EDR3), which is a consistent reduction of the public archival DECam data using the DESDM software pipeline \citep[][]{Morganson:2018} and was carried out as part of the \decade data processing campaign mentioned previously.  The primary advance in DELVE EDR3 relative to previous DELVE data releases is the generation of coadd images and catalogs, as well as multi-epoch data products and object fits that serve as the basis of high-quality shape measurements discussed here. The processing of individual DECam images, the coaddition of multiple DECam images, the detection of sources, and the measurement of their properties follow similar configurations to DES Data Release 2 \citep[DES DR2,][]{DES2021} and DES Y6 \citep{Bechtol:2025:Y6GOLD}. The DELVE EDR3 processing is also briefly described in \citet{Tan:2024}.

The DELVE EDR3 dataset is assembled from the region of the north Galactic cap that is observable by DECam. We select exposures that pass a set of quality criteria that are described in \citet{Tan:2024} and are also summarized below: 
\begin{itemize}
    \item Located in the northern Galactic cap, defined with 
    Galactic latitude $b > 10\degr$. The data contain images at sky-locations up to ${\rm Dec.} \lesssim 33\degr$, a limit imposed by the location of the Blanco telescope in the Southern Hemisphere,\vspace{5pt}
    \item Exposure time between $30 \leq t_{\rm exp} \leq 350$\,seconds, since the DESDM processing pipeline has not been extensively tested for coadding images with a broader range of exposure times,\vspace{5pt}
    \item PSF full-width half-max (FWHM) of $0 \arcsec < \text{FWHM} < 1.5 \arcsec$, \vspace{5pt}
    \item Good image quality \citep[see][]{Tan:2024} in addition to $t_{\rm eff} > 0.2$, where $t_{\rm eff}$ is a summary of the exposure quality \citep{Neilsen:2016}, \vspace{5pt}
    \item Good astrometric solutions, determined via comparisons with \textit{Gaia} DR2. See Section 2 in \citet{Tan:2024} for the precise comparison criteria,\vspace{5pt}
    \item Number of objects detected in each exposure is $< 7.5 \times 10^5$, to remove exposures with many spurious object detections caused by excessive electronic noise or poor background subtraction,\vspace{5pt}
    \item Images are further checked for quality by automated algorithms and human visual inspection.\vspace{5pt}
\end{itemize}

This results in 42,700 exposures across the $griz$ bands \citep{Tan:2024}. With these exposures, we produce the multi-epoch coadded images \citep{Morganson:2018}, \textsc{SourceExtractor} \citep{Bertin:1996:SrcExt} source detection catalogs, multi-epoch data files \citep[MEDs;][]{Jarvis2016}, and catalogs with the mutli-epoch, multi-object fits \citep[e.g.,][]{Sheldon:2014, Drlica-Wagner:2018}. In particular, we use the $r+i+z$ coadd images for object detection. The $g$ band is not used for detection, or for any downstream postprocessing such as shear measurement or redshift estimation, as is the case with DES Y3. The image coadds are produced in distinct rectangular tiles that have dimensions of $0.71\degr \times 0.71\degr$, where each dimension is covered by 10,000 pixels with a pixel scale of 0.263$\arcsec$. A coadd tile is built and processed only if it has at least one exposure in each of the four bands ($griz$) intersecting that region on the sky. The EDR3 region contains 16,895 coadd tiles covering a total area of $\approx\! 8,\!500$ deg$^2$.\footnote{While the $g$-band is not used in the lensing analysis, it is still a valuable input to a variety of science cases beyond lensing. Thus, we require 4-band coverage when making coadded images.} Note that this is prior to all additional selections needed to assemble the cosmology-ready galaxy sample. The latter covers $5,\!412\,\deg^2$ as mentioned previously. 

In the DESDM processing pipeline \citep{Morganson:2018}, a PSF model is fit to each individual CCD image using \textsc{PSFEx} \citep{Bertin2013} prior to image coaddition. These models are used in all object fits provided in the general-purpose DES ``Gold'' catalogs \citep{Drlica-Wagner:2018, Sevilla-Noarbe2021, Bechtol:2025:Y6GOLD}. These models were also used for weak lensing analyses in the DES Science Verification (SV) and Y1 datasets \citep{Jarvis2016, Zuntz2018}. 
However, more recent lensing analyses from DES Y3 and Y6 \citep[\citetalias{y3-shapecatalog},][]{Yamamoto2025} fit and use a different PSF model constructed via the ``PSFs in the Full Field-of-View'' software \citep[\textsc{Piff},][]{Jarvis2020}. In DES Y3 and Y6, the \textsc{Piff} models are generated in addition to the \textsc{PSFEx} ones, via a dedicated post-processing step. In \decade, we did not have the resources to post-process all single-epoch CCD images to generate \textsc{Piff} solutions for the image PSFs. Therefore, \textsc{PSFEx} serves as our PSF model.

The use of \textsc{PSFEx} rather than \textsc{Piff} introduces several differences in our PSF modeling compared to the latest DES Y6 models \citep{Schutt:2025:Y6PSF}. The most important of these differences between \textsc{PSFEx} and \textsc{Piff} are that the former uses a different star selection criteria compared to the latter, and the former also does not have any color-dependent modeling as present in the latter. Though, we note that the DES Y3 shape catalog \citepalias{y3-shapecatalog} used a color-independent PSF model from \textsc{Piff} for its shape measurements. Additionally, \textsc{Piff} models the PSF in sky coordinates, rather than image coordinates, which helps alleviate (but not remove) the impact of ``tree rings'' in the PSF solutions; see Figure~\ref{fig:psf_focalplane} and discussions therein. The systematic diagnostic tests in Section~\ref{sec:sys_tests} characterize our PSF model and show it has a small impact on the lensing data vector, while \citetalias{paper3} explicitly confirms that the final cosmological results are insensitive to any residual, additive contaminations from the PSF. Details of the \textsc{PSFEx} model choices are recorded in Appendix~\ref{sec:psfex}. 

\subsection{Base catalog selection}
\label{sec:base_sel}

Analogous to the ``Gold catalog'' in DES \citep{Drlica-Wagner:2018,Sevilla-Noarbe2021, Bechtol:2025:Y6GOLD}, we define a ``base catalog'' that serves as a starting point for any cosmology analysis on this data. This is obtained by performing the following selection cuts on the \textsc{SourceExtractor} catalog generated from the coadded images:

\begin{itemize}
    \item \textbf{\textsc{SourceExtractor} flags.} We select objects according to the following \textsc{SourceExtractor} flags: \texttt{FLAGS\_[GRIZ]<=3}, \texttt{IMAFLAGS\_ISO\_[GRIZ]==0}. This cut removes any objects that are clear artifacts.\vspace{5pt}

    \item \textbf{\textsc{Fitvd} flags.} All \textsc{SourceExtractor} objects are processed through the \textsc{Fitvd}\footnote{\url{https://github.com/esheldon/fitvd}} pipeline \citep{Bechtol:2025:Y6GOLD}, which jointly fits all exposure-level cutouts of the object in all bands. We remove any objects that have failed \textsc{Fitvd} measurements.\vspace{5pt} 
    
    \item \textbf{Foreground masking.} Following DES Y3, we mask a number of regions in the survey footprint. First we cutout circles around bright sources: 
    \begin{itemize}[label=$\circ$]
        \item Stars from 2MASS \citep{Skrutskie:2006:2Mass},
        \item Stars from \Gaia eDR3 \citep{Gaia:2021:eDR3},
        \item The Yale bright star catalog \citep{Hoffleit:1991:yalestars},
        \item A list of very bright stars from the BSCP5 catalog\footnote{These bright stars, some of which will also be included in the aforementioned catalogs, are masked separately from the rest. Their brightness necessitates larger masks because these sources cause scattered light artifacts in the DECam images.} \citep{Hoffleit:1991:BSCP5},
        \item Galaxies from the Hyperleda catalog\footnote{\url{http://leda.univ-lyon1.fr/}} \citep{Makarov:2014:Hyperleda}
        \item Globular clusters from the Harris catalog \citep[][2010 edition]{Harris:1997:GlobularClusters} 
    \end{itemize}
     We additionally remove areas of the sky with high stellar density, defined by the number density of \Gaia objects (with $G < 21$\,mag) exceeding $5$ arcmin$^2$, and also areas with high interstellar extinction, defined by $E(B - V) > 0.2$ according to the extinction map of \citet{Schlegel:1998:Dust}. These cuts ensure that the stellar density and dust extinction in the EDR3 footprint have a similar maximum bound as seen in the DES Y3 region of the sky. The final foreground mask removes roughly 30\% of objects and about 2,000 deg$^2$. \vspace{5pt}

    \item \textbf{Other masks.} We finally remove a small number of objects in regions that either have a statistically significant surplus of unphysical colors\footnote{These objects are defined with $\texttt{SNR} > 10$, and having colors \textit{outside} the range $-1 < X < 4$ for any of the colors $X \in \{g-r, r-i, i-z\}$. We then identify regions of the sky (partitioned into \textsc{HealPix} pixels of $\texttt{NSIDE} = 256$) with at least 10 selected objects and mask the area in these pixels. This cut is separate from the colors cuts presented in Section~\ref{sec:shear}, where the latter are motivated by details in the redshift estimation procedure.} (indicating failures in zeropoint calibration or astrometry) or are located in areas that are very disjoint from the main contiguous footprint. This removes $<1\%$ of objects that pass all the other cuts above.
\end{itemize}

After all cuts above, we are left with 470,812,637 objects covering an area of $\approx\! 5750$ deg$^{2}$. While we broadly follow the procedure used in DES Y3 \citep{Sevilla-Noarbe2021}, we do not include flag bit 32 and 64 from their Table 3, which correspond to bright blue artifacts in the images and bright objects with unphysical colors, respectively. These objects are extremely rare --- they constituted $<0.3\%$ of the base catalog in DES Y3 \citep[][see their Table 3]{Sevilla-Noarbe2021} --- and most do not survive our shape catalog selection described in the next section. For example, the DES Y3 color cuts corresponding to flag bit 64, which are defined as $-1 < X < 4$ with $X = a - b$ being the magnitudes in two different bands, are nearly the same as the color cuts placed on the galaxy shape catalog (see Section~\ref{sec:selection}). We verified, using the DES Y3 catalog and flags, that such bright objects and unphysical color objects only constitute $0.01\%$ and $0.003\%$ of the catalog after the sample selection defined above.

\subsection{Differences relative to DES}\label{sec:diff2DES}

The image processing used in deriving the \decade shape catalog is heavily based on that of DES. However, there are three areas where our dataset and processing \textit{differ} from DES Y3, which we describe below for reference.

\textbf{Survey inhomogenity.} DES observed 5,000\,deg$^2$ of the sky with ten 90-second exposures in $griz$ and an equivalent depth of ten 45-second exposures in $Y$. It used real-time data quality assessments to guide observing decisions (\ie telescope time during poor observing conditions was allocated to supernova observations; \citealt{Neilsen:2019}). Our data, on the other hand, is significantly less homogeneous. The DECam survey programs covering large areas of the sky each followed a variety of strategies for optimizing homogeneity \citep[e.g.,][]{Dey:2019,Drlica-Wagner:2021}, while the presence of additional survey and standard programs covering smaller fractions of the sky necessarily lead to additional inhomogeneity in our final dataset. We refer readers to \citet{Tan:2024} for a description of the exposure selection process used to reduce some of these inhomogeneities prior to image processing.

\textbf{Airmass distribution.} The DECam optics were not designed to remove differential chromatic refraction (DCR) from the atmosphere \citep[e.g.,][see their Section 4.7]{Bernstein:2017:Astrometry}. For this reason, the DES wide-area footprint and corresponding observing strategy are defined to have an airmass of ${\rm sec}(z) \leq 1.4$, where $z$ is the zenith angle. The northern Galactic cap surveyed in this work naturally extends to higher airmass --- $25\%$ of our catalog objects are in regions with ${\rm sec}(z)>1.4$ and $4\%$ are in regions with ${\rm sec}(z)>1.8$ --- which then results in a residual point spread function (PSF) modeling error due to DCR.\footnote{The PSF size and ellipticity are color-dependent due to the DCR effect. Across the wavelength range spanned by each filter band, stars and galaxies can have slightly different spectral energy distributions (SEDs); stars are usually bluer than galaxies. Thus, when we use stars to build a PSF model and apply the model to galaxies when estimating shapes, we introduce a small error that is associated both with the difference in the SED of stars and galaxies, \textit{and} with the amplitude of the DCR effect. This amplitude is greater at high airmass.} We do not detect a significant correlation between measured shear and DCR as a result of this effect (see Section~\ref{sec:SysMaps} and the tests in \citetalias{paper3}).
    
\textbf{PSF model.} As mentioned previously, the PSF model algorithm used for shape measurements in this catalog, \textsc{PSFEx} \citep{Bertin2013}, is different from the \textsc{Piff} \citep{Jarvis2020} algorithm used in the most recent DES lensing analyses \citep[\citetalias{y3-shapecatalog},][]{Yamamoto2025}. We extensively test the impact of the different PSF model used in our work and find no significant impact of the PSF systematics on the shape catalog (see Section~\ref{sec:sys_tests} and  also the tests in \citetalias{paper3}). We provide additional details on our PSF modeling approach, and on its differences from the DES Y3/Y6 analysis choices, in Appendix~\ref{sec:psfex}.

These differences between \decade and DES provide us with the opportunity to study the limits on image/data quality at which various algorithms still function without introducing substantial systematic biases. These results could have implications for the next generation of datasets, such as LSST, where a significant amount of additional image data could be incorporated into the analyses if we are able to extend lensing datasets to regions with higher airmass and less homogeneous coverage. See \citet{Bianco:2022:FootprintLSST} for an example of the proposed LSST footprint.

\section{shape catalog}
\label{sec:shear}

A weak lensing analysis requires estimating some tracer of the lensing shear field, often taken to be the galaxy shapes. Typically, we use measurements of the two-component ellipticity $\boldsymbol{e}$, with $\boldsymbol{e}=e_{1} + i e_{2}$. Various methods have been used to measure these ellipticities, though a precise measurement is made challenging due to the presence of noise, pixel masks, the PSF, selection effects, etc. of our observations. In the presence of external cosmological shear $\boldsymbol{\gamma}=\gamma_{1} + i \gamma_{2}$, the observed ellipticities change, and the resulting change in a particular shape estimate defines the \textit{response}, typically denoted $R$. In the case when $\boldsymbol{e}$ is the shape estimate, we can schematically write $R=\delta \boldsymbol{e}/\delta \boldsymbol{\gamma}$. Having an unbiased measure of $R$ is essential to achieving unbiased cosmological constraints. Over the past decade, there has been a rich literature devoted to developing algorithms that can robustly measure $R$, a process sometimes referred to as ``shear calibration'' \citep[see e.g.,][for an overview of the various methods]{Mandelbaum2018}. Of these, we use the \textsc{Metacalibration} algorithm \citep{Sheldon2017, Huff2017}. In this work, all shear-related estimates are performed using image data from only the $riz$ bands; we discard the $g$-band data, following the DES Y3 findings of  significant PSF modeling uncertainties in that band \citep{Jarvis2020}. Similar to the approach in DES Y3 and DES Y6 \citep{y3-shapecatalog, Maccrann2022ImSim, Yamamoto2025}, the estimates are performed by jointly fitting all single-epoch cutouts of a galaxy (i.e. the observations of galaxies in individual CCD images) across the $riz$ bands.

\subsection{\textsc{Metacalibration}}\label{sec:metacal}

\textsc{Metacalibration} \citep{Sheldon2017, Huff2017} is a shear calibration method and is the fiducial method used in \citetalias{y3-shapecatalog}. We refer the reader to the original papers for details of the method. Here we only sketch the conceptual idea behind the method and introduce the equations that we will refer to later in this work. 

The key idea of \textsc{Metacalibration} is that we can estimate the response, $R$, by artificially shearing the data at the pixel level. Particularly, for each galaxy, the method requires five versions of the same image: (no shear, $1+, 1-, 2+, 2-$). Aside from the ``no shear'' version, which is the original, the other four are obtained by shearing the images by a small amount ($\Delta \gamma_{i\pm}=\pm 0.01,\; i=1,2$) positively/negatively in the two sky coordinates 1 and 2. The same shape measurement code is run on all five versions of the image cutouts of this galaxy, resulting in five versions of the galaxy's ellipticity measurement. In this work, our shape measurements are based on \textsc{NGMix} \citep{Sheldon:2014} and fit galaxy images using an elliptical Gaussian convolved with the PSF model. In addition to the ellipticities, we also measure all the other object properties (such as photometry) in all five variants associated with the galaxy.

We can now calculate two response factors from the five measurements detailed above. The first response factor $\boldsymbol{R_{\gamma}}$ is the response of the ellipticity estimate to external shear, as discussed earlier, and it can be calculated via
\begin{equation}\label{eqn:response_reg}
[\boldsymbol{R_{\gamma}}]_{ij} = \frac{e_{i}^{\,j+} - e_{i}^{\,j-}}{2 \times 0.01}; \; i,j=1,2.
\end{equation}
We note that $\boldsymbol{R_{\gamma}}$ can be calculated for every galaxy; though for a single galaxy the estimate is generally dominated by noise. When calibrating summary statistics used for cosmological inference, we follow DES analyses and only use the average response of the relevant sample, in which case the response estimate is no longer noise dominated.

The second response factor, denoted $\boldsymbol{R}_{S}$, is related to how external shear also affects the selection of the galaxy sample \citep[][see their Equations 12-14]{Sheldon2017}. This is because objects can pass/fail a given selection when sheared. ``Selection'' here refers to the cuts we place on the sample, using quantities such as galaxy size and signal-to-noise. We discuss the full list of selections in Section~\ref{sec:selection}. This selection response is calculated using the five versions of the shape measurements. In particular, we perform the same selection using the $i+$ and $i-$ versions of \eg galaxy size, signal-to-noise, and any other \textsc{Metacalibration} quantity used in the selection function. The selection response is then   
\begin{equation}
\label{eqn:response_selection}
    \langle \boldsymbol{R}_{S}\rangle_{ij} \approx \frac{\langle e_{i}\rangle^{S_{i+}} - \langle e_{i}\rangle^{S_{i-}} }{\Delta \gamma_{j}},
\end{equation}
where $S$ denotes the full set of selections applied to the catalogs, and $S_{i+}$ denotes that the selections were performed using measurements from the $i+$ sheared images as extracted from \textsc{Metacalibration}. Note that $\boldsymbol{R}_{S}$ can only be calculated for an ensemble since the selection effect is fundamentally a response of the sample; the selection $S_{i+}$ and $S_{i-}$ is not guaranteed to return the same galaxies, and that is precisely the effect being calibrated. We briefly mention that the DES Y6 catalog \citep{Yamamoto2025} uses the \textsc{Metadetection} algorithm \citep{Sheldon2020}, which also accounts for a detection-based response in addition to the terms above (since detection is also formally part of the selection term). For this work, which uses \textsc{Metacalibration}-based estimates, we follow DES Y3 in calibrating such detection-based effects using suites of image simulations, as undertaken in DES Y3 \citet{Maccrann2022ImSim} and described in Section~\ref{sec:calibration_bias} below.

Finally, the full response factor for a sample of galaxies is the sum of the two response factors above, or
\begin{equation}\label{eqn:combined_response}
\langle \boldsymbol{R}\rangle = \langle \boldsymbol{R}_{\gamma}\rangle + \langle \boldsymbol{R}_{S}\rangle.
\end{equation}
In practice, we take the $\langle \boldsymbol{R} \rangle$ matrix to be diagonal, since the off-diagonal terms are negligible, and also take the mean of the two diagonal elements as the single response factor when performing the cosmic shear measurements. Our implementation, including all assumptions therein, exactly follow the procedure verified and chosen in \citetalias{y3-shapecatalog}.

\subsection{Shear sample selection}
\label{sec:selection}

\begin{figure}
    \centering  
    \includegraphics[width=0.49\textwidth]{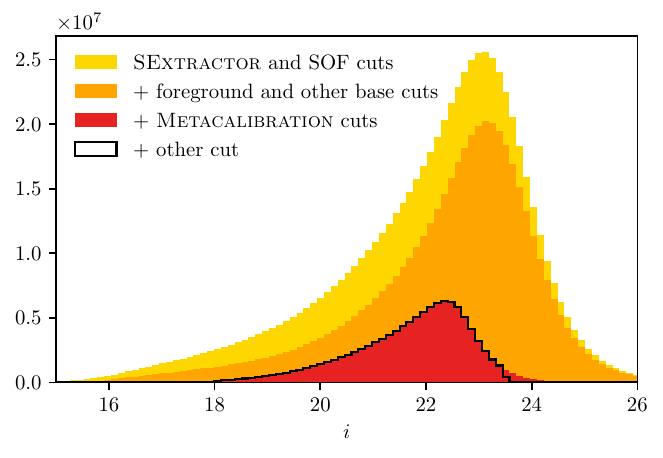}
    \caption{The $i$ band magnitude distribution as a function of different selection cuts applied to derive the final shape catalog. The size of the final sample (black) is approximately 16\% the size of the sample with only \textsc{SourceExtractor} and \textsc{Fitvd} cuts (gold).}
    \label{fig:mag_dist_cuts}
\end{figure}

\begin{figure*}
    \includegraphics[width = 2\columnwidth]{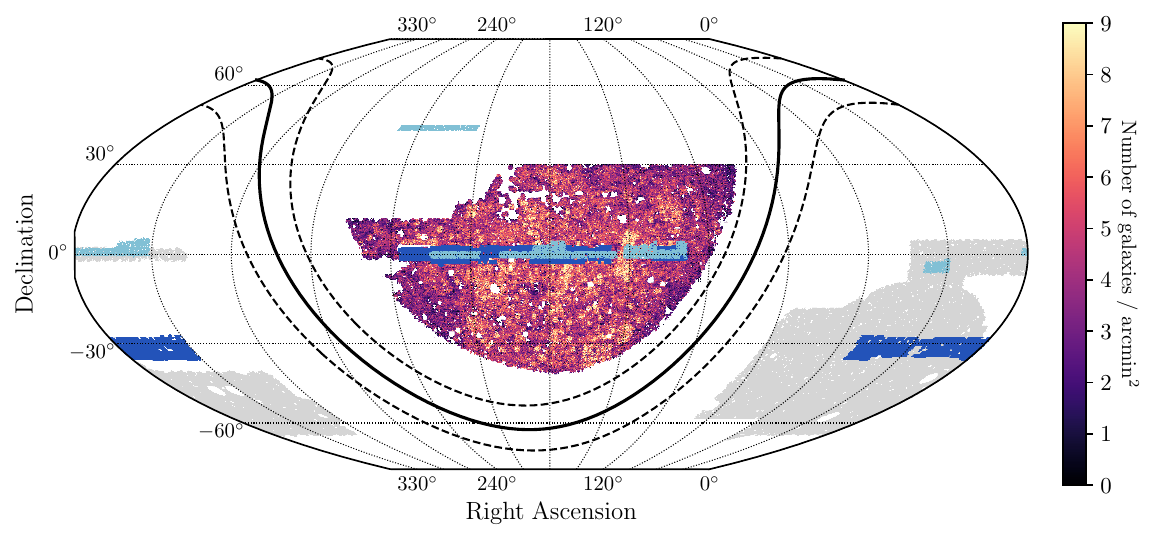}
    \caption{Map of the number density of galaxies in our shape catalog. We also show the footprint of three other Stage-III lensing surveys: DES Y3 (grey), KiDS-1000 (dark blue), HSC Y3 (light blue). The dashed lines indicating the Galactic plane with $b = \pm 10\degr$. Our catalog consists of 107 million galaxies covering an area of 5,412\,deg$^2$. The distance between our footprint and the Galactic plane is a result of our stellar density mask. This mask is also asymmetric due to the presence of the Galactic bulge.} 
    \label{fig:footprint}
\end{figure*}

\begin{figure}
    \centering  \includegraphics[width=0.48\textwidth]{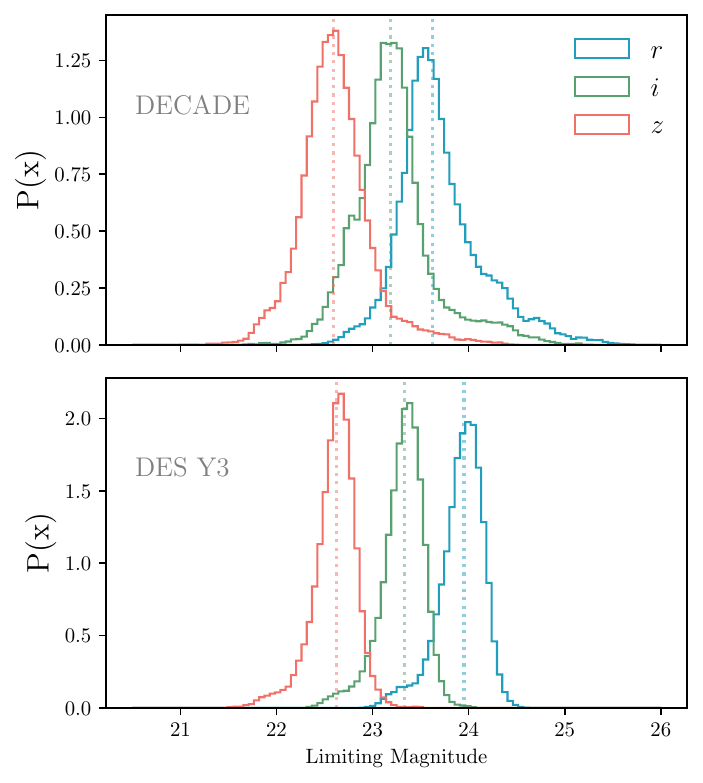}
    \caption{Distribution of the limiting magnitude in the $riz$ bands across the footprint for our shape catalog (top) and the DES Y3 catalog (bottom). The median limiting magnitudes, shown as vertical dotted lines, are $r = 23.63 (23.95)$, $i = 23.18 (23.34)$, and $z = 22.59 (22.63)$ for \decade (DES). The \decade data has a similar median magnitude as DES Y3, but a considerably wider spread. Around 10\% of our area has a higher magnitude limit than the 99\% maximum value in DES Y3 (for each band). This map only shows spatial variations in the imaging data quality, and not in the source-galaxy number densities, which is cut at brighter magnitudes and therefore shows less variation (see Figure 1 in \citetalias{paper3}).}
    \label{fig:maglim}
\end{figure}

Starting from the raw catalog containing all \textsc{SourceExtractor} detections, we closely follow \citetalias{y3-shapecatalog} and place the following selection cuts to form the final shape catalog:

\begin{itemize}
    \item The base selection described in Section~\ref{sec:base_sel}\vspace{5pt}
    
    \item Have a corresponding measurement in the \textsc{Metacalibration} catalog\footnote{Not every \textsc{SourceExtractor} detection has a \textsc{Metacalibration} measurement, as the latter explicitly removes any galaxy for which we do not have simultaneous coverage across the $riz$ bands. Thus, this selection implicitly contains such a cut as well.} and pass the \textsc{Metacalibration} cuts,
    \begin{align}
    &{\rm 10<SNR <1000;}\notag \\  
    &{\rm T/T_{psf}>0.5;} \notag \\ 
    &{\rm T<10;} \notag \\
    &{\rm NOT( (T>2) \; and \;  (SNR<30)).} 
    \end{align}
    Here $T$ and $\rm SNR$ are the size (in $\rm arcsec^2$) and signal-to-noise estimates for the galaxy, whereas $\rm T_{psf}$ is the size estimate of the PSF. All these quantities are outputs from \textsc{Metacalibration}.\vspace{5pt}

    \item Remove galaxies that (i) are very faint/bright in either of the three bands, and (ii) exhibit extreme colors that are unexpected for a galaxy sample, as both considerations help improve the photometric redshift uncertainties and calibration of the catalog \citep[][see their Section 3.1]{Myles:2021:DESY3}
    \begin{align}
     15 & < r < 26 \nonumber\\
     18 & < i < 23.5 \nonumber\\
     15 & < z < 26 \nonumber\\
     -1.5 & < r - i < 4 \nonumber\\
     -1.5 & < i - z < 4.
    \end{align}\vspace{5pt}

    \item Remove binary stars (or close stellar pairs), which can appear as highly elliptical sources due to blending and due to limited resolution in our images. These are discarded by selecting all objects with $\sqrt{e_1^2 + e_2^2} > 0.8$, and then removing those with
    \begin{equation}
    \log_{10}(\frac{T}{\rm arcsec^2}) > \frac{22.25-r}{3.5}.
    \end{equation}
    This cut was motivated by visual checks on the DES Y3 data \citepalias{y3-shapecatalog}. We have verified the presence of these objects in our catalog as well and therefore replicate these cuts in our analysis.\vspace{5pt}

    \item Finally, when performing a tomographic cosmic shear analysis, we place these galaxies into tomographic bins according to their redshift estimates. We define four tomographic bins roughly in the redshift range of $0<z<2$ \citepalias[see details in][]{paper2}. Note that no galaxies are discarded during this selection. \vspace{5pt}
\end{itemize}

All magnitudes in the above selection use \textsc{Metacalibration} fluxes rather than the fluxes estimated by the \textsc{Fitvd} model fit described in Section \ref{sec:base_sel} -- the use of the former photometry is required to accurately employ the \textsc{Metacalibration} response formula (Equation~\ref{eqn:response_selection}) and derive the selection bias. All color/magnitude cuts are made on \textit{de-reddened} photometric estimates, where the de-reddening correction $A_b \times E(B-V)$ in band $b$ uses the $E(B-V)$ map of \citet{Schlegel:1998:Dust} with the filter-dependent factors $A_b$ as calculated and presented in DES DR1 \citep[][see their Section 4.2]{DES2018}. The reddening coefficients are 2.140, 1.569, 1.196 for the $riz$ bands.

The final selection function results in a catalog of 107,371,478 galaxies, covering $5,\!412 \deg^2$ of the sky. Figure~\ref{fig:mag_dist_cuts} and Table~\ref{tab:selection} show the changes in the sample size as a function of the different cuts. Figure~\ref{fig:footprint} shows the number density of the galaxies in the final shape catalog and also plots footprints from other external datasets for comparison. The number density of our catalog is generally higher in the center of the footprint and lower on the eastern/western edges (where stellar density increases due to proximity to the Galactic plane), and some fluctuations in this number density exhibit sizes corresponding to the DECam focal plane. The latter fluctuations are almost exclusively in regions that are deeper than the median depth of the \decade data (see Figure \ref{fig:SP_maps}). This is expected given the origin of our dataset, which is an amalgamation of community DECam data used for various science goals. In \citetalias{paper3} (see their Section 6), we perform numerous tests that indicate the impact of such variations on our cosmology constraints is below the precision of the dataset.

In order to make a direct comparison between the imaging data of the \decade and DES datasets, we show the 10$\sigma$ limiting magnitude for photometry computed with $2\arcsec$ apertures. Figure~\ref{fig:maglim} shows the distribution of these limiting magnitudes over the footprint spanned by our shape catalog and that by the DES Y3 sample. The median limiting magnitude of our data in each band is $r= 23.6$, $i= 23.2$, $z = 22.6$. These numbers can be directly compared with the ``Median Coadd Magnitude Limit'' in Table 2 of 
\cite{Sevilla-Noarbe2021}, where they find $r = 24.0$, $i = 23.3$, $z = 22.6$ -- our imaging data is slightly shallower than DES Y3 in $r,i$, and similar to it in $z$. The image depth varies significantly more in \decade than DES Y3, as is expected. In fact, around 10\% of the \decade footprint is \textit{deeper} than the 99\% value found in the DES Y3 distribution. We stress, however, that this only characterizes the image depth and not number density of the source galaxy catalog. Figure 1 in \citetalias{paper3} shows that the spatial variations in our source galaxy number density have a \textit{similar amplitude} to that seen in DES Y3. The magnitude cuts in the selection function described above, which were motivated by redshift precision considerations, reduce inhomogenieties in the source galaxy number density (as caused by the inhomogeneous image depth/quality).

\begin{table}
    \centering
    \begin{tabular}{l|l|l}
     \hline
     Selection type & Number (\%)  & Cumulative number (\%) \\
     \hline
     \textsc{SrcExt} + \textsc{Fitvd} & 671,290,620 (100\%) & 671,290,620 (100\%)\\
     Foreground + base & 470,812,637 (70.14\%) & 470,812,637 (70.14\%)\\
     \textsc{Metacalibration} & 140,432,055 (20.92\%) & 111,400,589 (16.59\%) \\
     Other & 480,429,730 (71.57\%)& 107,371,478 (15.99\%)\\
    \hline
    \end{tabular}
    \caption{Number of galaxies as a function of the different selection cuts applied to form the shape catalog. The first column shows the number of galaxies (and percentage relative to the first row) that pass a single type of cut, while the second column shows the cumulative numbers from combining the cuts sequentially, where the sequence is ordered from top to bottom.}
    \label{tab:selection}
\end{table}

\begin{figure*}
    \centering  \includegraphics[width=0.35\textwidth]{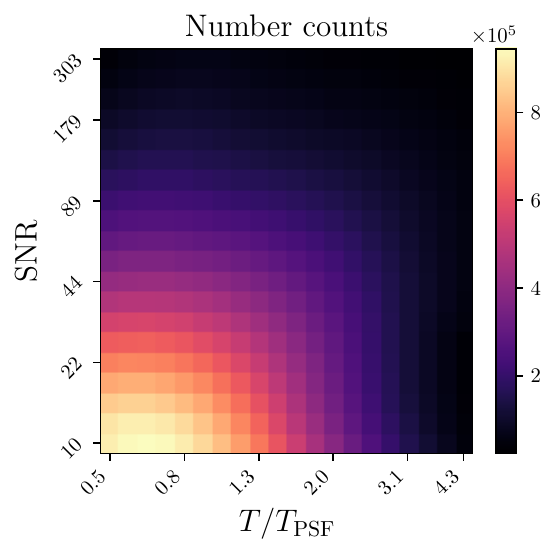}
    \includegraphics[width=0.35\textwidth]{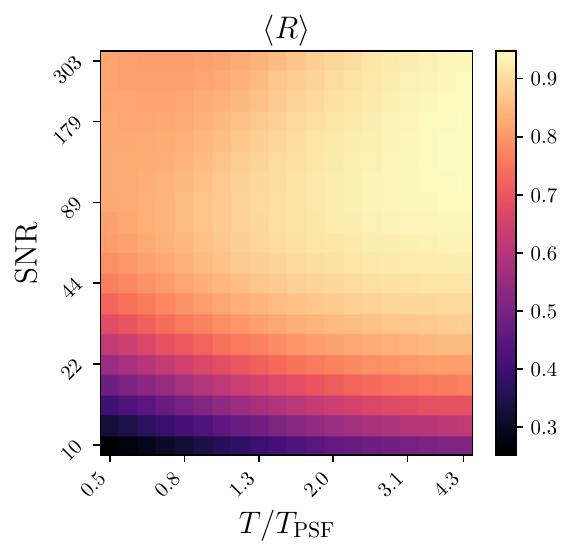}
    \includegraphics[width=0.37\textwidth]{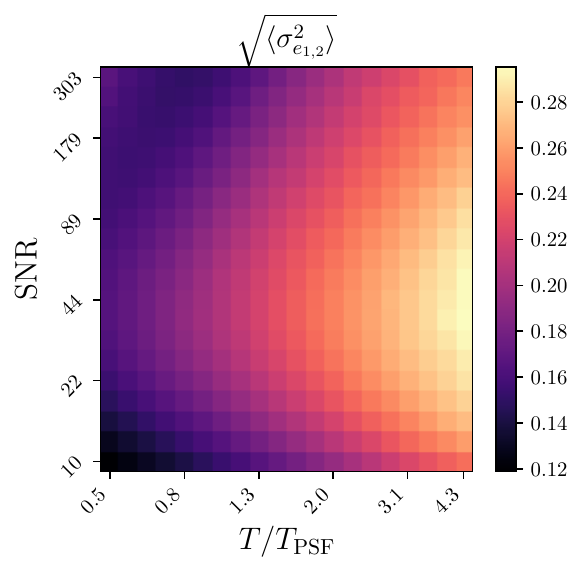}
    \includegraphics[width=0.35\textwidth]{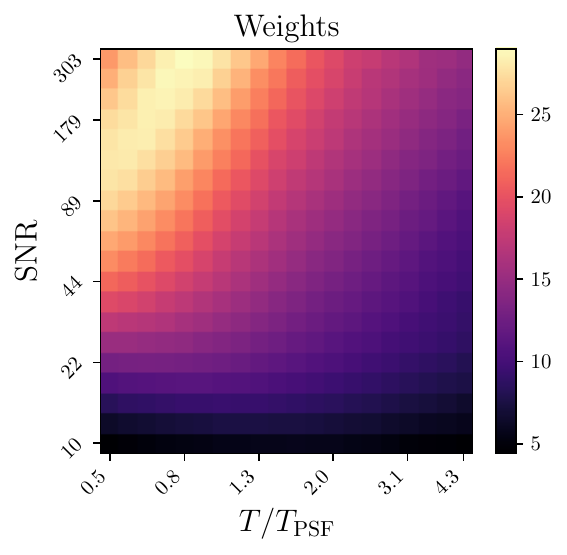}
    \caption{Number counts (upper left), shear response $\langle R \rangle$ (upper right), shape noise $\sqrt{\sigma_{e}^2}$ (lower left) and shear weights (lower right) as a function of size ratio $(T/T_{\rm PSF})$ and signal-to-noise (SNR) of the galaxy. For each galaxy in the final shape catalog, a weight is assigned according to the lower right panel. Including this weight enhances the signal-to-noise of any statistic computed with the shape catalog.}
    \label{fig:weights}
\end{figure*}

\subsection{Weights}\label{sec:weights}

In this work, each galaxy is assigned a weight that optimizes the signal-to-noise of statistics computed with the galaxy shapes. That is, galaxies with noisier shear measurements will be down-weighted and vice versa. Following \citetalias{y3-shapecatalog}, the weight of each galaxy is assigned according to their signal-to-noise\footnote{For clarity, we reproduce footnote 8 from \citetalias{y3-shapecatalog} on the definition of this signal-to-noise: ``we define the signal-to-noise ratio (S/N) as $\text{S/N} = \left(\sum_p m_p I_p / \sigma_p^2\right) / \left(\sum_p m_p^2 / \sigma_p^2\right)^{1/2}$ where the sum runs over the pixel \( p \), \( m_p \) is the best fit model for the galaxy, \( I_p \) the measured pixel value, and \( \sigma_p \) the estimated pixel variance.''} and size, as estimated in \textsc{Metacalibration}. We first calculate two quantities $(\sigma_{e}^2)_{jk}$, $\langle R \rangle _{jk}$ for galaxies binned into size ratio bin $j$ and signal-to-noise bin $k$. Here, the size ratio is the size of the galaxy to the PSF, $(T/T_{\rm PSF})$. Following \citetalias{y3-shapecatalog}, we choose 20 logarithmic bins in size ratio from $0.5 < (T/T_{\rm PSF}) < 5$ and 20 logarithmic in signal-to-noise from $10 <  {\rm SNR} < 330$. This range includes $97.5\%$ of all the source galaxies. Remaining galaxies with large SNR or $(T/T_{\rm PSF})$ are subsumed into the last bin of the respective quantity.

The quantity $(\sigma_{e}^2)_{jk}$ is the variance in the single-component ellipticity within a cell $jk$ of the 2D grid, defined as
\begin{equation}
(\sigma_{e}^2)_{jk} = \frac{1}{2N_{jk}}\sum_{i}^{N_{jk}} (e_{1,i}^2 + e_{2,i}^2),
\end{equation}
where $N_{jk}$ is the number of galaxies in bin $jk$, while $e_{1}$ and $e_{2}$ are the single-component, uncalibrated ellipticities from galaxies in bin $jk$. $\langle R \rangle_{jk}$ is the mean, unweighted shear response in bin $jk$ as calculated using Equation~\ref{eqn:combined_response}.

Each bin is then assigned the weight 
\begin{equation}
w_{jk} = \frac{\langle R \rangle_{jk}^2}{(\sigma_{e}^2)_{jk}}
\label{eq:weights}
\end{equation}
and all galaxies in cell $jk$ are assigned weight $w_{jk}$. For galaxies with sizes and/or signal-to-noise values outside the grid defined above, we assign weights corresponding to the closest grid point. Figure~\ref{fig:weights} shows the number counts, $\sigma_{e}^2$, $\langle R \rangle$ and $w$, as a function of size ratio and signal-to-noise. Compared with Figure 4 of \citetalias{y3-shapecatalog}, we see fairly similar patterns in the number counts, $\langle R \rangle$ and weights. There are some minor difference in the shape noise distribution, but these are \textit{a priori} expected due to the difference in image quality between our data and DES Y3. Thus, they are not signs of any underlying issue in our data.

\begin{table*}
    \centering
    \begin{tabular}{c|c|c|c|c|c|c|c|c|c|c|c|c|c}
     \hline
     & $n$ & $R_{\gamma, 1}$ & $R_{S, 1}$ & $R_{\rm tot, 1}$ & $R_{\gamma, 2}$ & $R_{S, 2}$ & $R_{\rm tot, 2}$ & $n_{\rm eff, C13}$ & $\sigma_{e, {\rm C13}}$ &$n_{\rm eff, H12}$ & $\sigma_{e, {\rm H12}}$ & $\langle \gamma_{1}\rangle$  & $\langle \gamma_{2}\rangle$ \\
     \hline
    Bin 1 & 1.396 & 0.850 & -0.014 & 0.836 & 0.851 & -0.014 & 0.837 & 1.229 & 0.232 & 1.239 & 0.233 & 0.00012 & $-0.00059$ \\
    Bin 2 & 1.371 & 0.761 & 0.009 & 0.771 & 0.762 & 0.009 & 0.771 & 1.130 & 0.257 & 1.150 & 0.259 & 0.00026 & $-0.00078$ \\
    Bin 3 & 1.375 & 0.721 & 0.018 & 0.740 & 0.722 & 0.020 & 0.742 & 1.134 & 0.244 & 1.169 & 0.248 & 0.00026 & $-0.00090$ \\
    Bin 4 & 1.369 & 0.592 & 0.028 & 0.620 & 0.593 & 0.028 & 0.621 & 1.054 & 0.276 & 1.153 & 0.289 & 0.00051 & $-0.00108$ \\
    Full sample & 5.511 & 0.748 & 0.008 & 0.756 & 0.750 & 0.008 & 0.757 & 4.474 & 0.250 & 4.586 & 0.254 & 0.00025 & $-0.00079$ \\
        \hline
    \end{tabular}
    \caption{The number density ($n$), different components of the shear response ($R_{\gamma/S/{\rm tot}, 1/2}$), effective number density of source galaxies ($n_{\rm eff}$) and shape noise ($\sigma_{e}$) in the H12 and C13 definitions, and the mean weighted shear ($\langle \gamma_{1,2} \rangle$), all computed for each of the tomographic bins as well as the full non-tomographic sample. The number densities are calculated with an area of $5,\!412 \deg^2$, and are presented in units of 1/arcmin$^2$.}
    \label{tab:neff_w2}
\end{table*}

\subsection{Response, number counts and shape noise}

Given the weights described above, we can estimate the effective number density of source galaxies, $n_{\rm eff}$, as well as the associated shape noise, $\sigma_{e}$, of our sample. These numbers, together with the area of our footprint $A$, provide a first-order estimate for the statistical power of this dataset. Following previous work, we calculate two versions of $n_{\rm eff}$ and $\sigma_{e}$ based on \cite{Heymans2012} and \cite{Chang2013} for easy comparison with other datasets.

First, following \citet{Heymans2012}, we define

\begin{align}
    n_{\rm eff, H12} & = \frac{1}{A} \frac{(\sum w_i)^2}{\sum w_{i}^2}\\[5pt]
    \sigma_{e, {\rm H12}}^2 & = \frac{1}{2} \left[ \frac{\sum(w_i e_{1,i})^2}{(\sum w_i)^2} + \frac{\sum(w_i e_{2,i})^2}{(\sum w_i)^2} \right] \left[ \frac{(\sum w_i)^2}{\sum w_{i}^2} \right]
\end{align}

Second, following \citet{Chang2013}, we define

\begin{align}
n_{\rm eff, C13} & = \frac{1}{A} \frac{\sigma_{e, {\rm C13}}^2 (\sum w_{i}R)^2}{\sum w_{i}^2 \bigg(R^2 {\sigma_{e, {\rm C13}}^2 + \frac{1}{2} (\sigma^2_{{\rm m}, 1, i} + \sigma^2_{{\rm m}, 2, i})\bigg)}}\\[10pt]
\sigma_{e, {\rm C13}}^2 & = \frac{1}{2} \frac{ \sum w_{i}^2 (e^{2}_{1,i}+e^{2}_{2,i}-\sigma^2_{{\rm m},1 , i} - \sigma^2_{{\rm m},2 , i}) }{ \sum w_{i}^2 R^2},
\end{align}
where $\sigma^2_{{\rm m}, 1, i}$ and $\sigma^2_{{\rm m}, 2, i}$ are the measurement noise estimated via \textsc{Metacalibration}. 

Table~\ref{tab:neff_w2} summarizes the final $n_{\rm eff}$ and $\sigma_{e}$ values for the two definitions and for both the full sample and also the sample split in four tomographic bins. In \citetalias{paper3} and \citetalias{paper4}, we show that the \decade shape catalog has similar statistical precision to the DES Y3 shape catalog. Table~\ref{tab:neff_w2} also includes the weighted, calibrated mean shear in each bin. The mean $\gamma_2$ in some bins (particularly bin 3 and 4) is relatively high in our data. This mean shear is subtracted following \citetalias{y3-shapecatalog} and \citet{Yamamoto2025}. In the tomographic analyses performed in the other works of this series \citepalias{paper2, paper3, paper4}, we subtract a mean shear per tomographic bin. We note that a non-zero mean shear has also been found in the aforementioned DES works but its origin is unclear. In \citetalias{paper3} (see their Section 6) we analyze forty-six different half-splits of the shape catalog, and find they return statistically consistent cosmology constraints. The tomographic bin-dependent mean shear can vary under each split (since a split alters the selection of the shape catalog), with the $95\%$ range across splits roughly spanning $\delta \langle\gamma\rangle \approx 5 \times 10^{-4}$ for the two shear components and the different tomographic bins.

\begin{figure*}
    \centering
   \includegraphics[width=2\columnwidth]{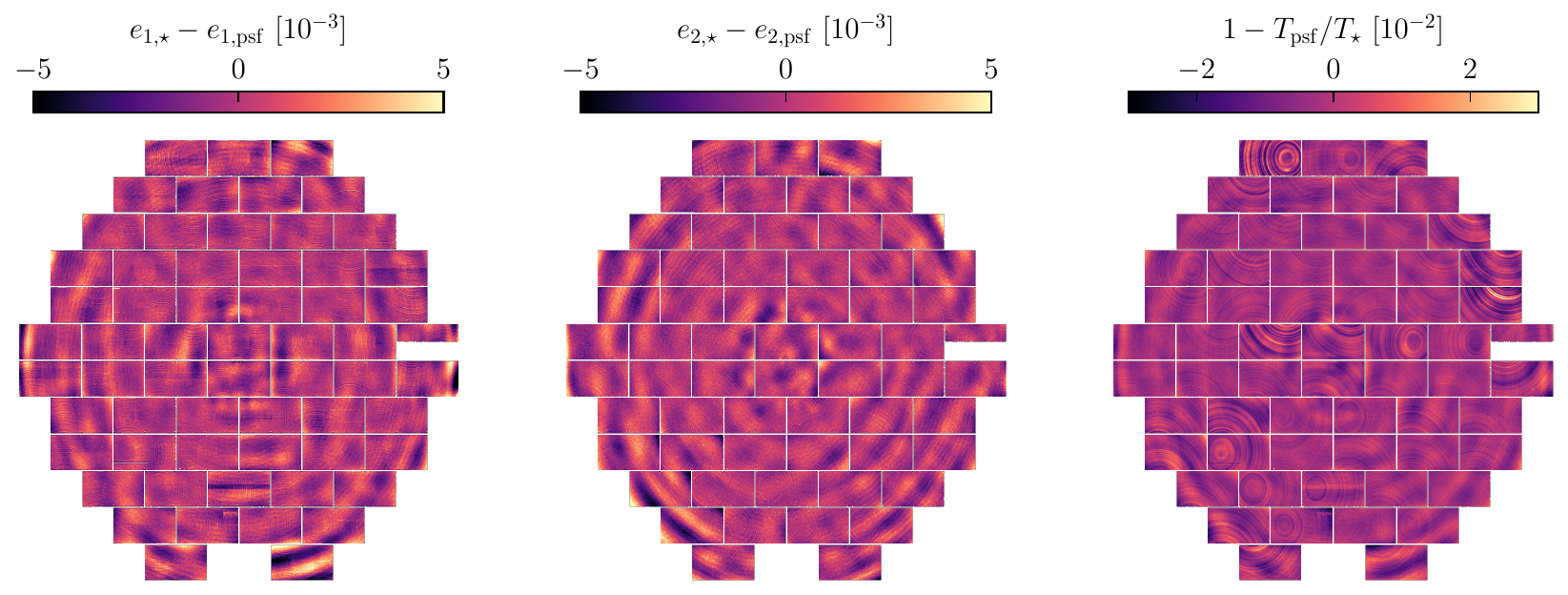}
    \caption{The mean PSF ellipticity error (left, middle) and size error (right) across the DECam focal plane. The large-scale, correlated modes in the ellipticity errors are from interpolation errors in the PSF model and follows those observed in DES \citep{Jarvis2020}. The size errors show small-scale features that are localized to individual CCDs, and these are ``tree rings'' arising from impurities in the silicon used in the sensors \citep{Estrada:2010:Treerings, Plazas:2014:TreeRings, Plazas:2014:TreeRings2}.}
    \label{fig:psf_focalplane}
\end{figure*}

\section{Empirical shear tests}
\label{sec:sys_tests}

Following the convention in the community \citep{Jarvis2016, Zuntz2018, y3-shapecatalog, Giblin2021,Li2022, Yamamoto2025}, we perform a number of tests to ensure our shape catalog is not impacted by systematic biases associated with instrumental or observational effects. These are primarily null tests that verify the quality of (or reveal oddities in) our data through analyses that use only the data itself (\ie without any simulations). Such tests do not isolate the exact origin of any detected oddities, but finding a statistically significant signal in any null test is a sign of concern and necessitates further investigation into the data. These tests are especially important in our case, as we are working with a somewhat unconventional weak lensing dataset where new effects could be relevant.

One of the primary systematic sources that can generate spurious shear is the imperfect modeling of the PSF. This is of particular concern because the PSF can contain spatial correlations which would then mimic the cosmological signal. Furthermore, we use a different algorithm (\textsc{PSFEx}) for the PSF modeling compared to that used in the most recent DES analyses \citep[\textsc{Piff},][]{Jarvis2020}, so these tests check that the quality of our PSF model is still sufficient for cosmological inference of this data. Tests associated directly with the PSFs are described in Sections~\ref{sec:psf_map}, \ref{sec:bf}, \ref{sec:psf_color}, part of \ref{sec:shearVsmcal}, and \ref{sec:rowe}. The PSF star sample used for these tests is detailed in Appendix~\ref{appx:PSFStarSample}.

In Section~\ref{sec:gammat_center}, we describe other tests that could be indirectly related to the PSF modeling systematics as well as other spatially varying systematic effects in the instrument and data. In Section~\ref{sec:SysMaps} we also test potential contamination in the shear measurements associated with observing conditions of the imaging data, and in Section~\ref{sec:b-modes} we quantify the $B$-mode contribution to our shape catalog.

For tests involving the spatial correlation of PSF ellipticities --- the Rowe/Tau stats (Section~\ref{sec:rowe}) and tangential shear (Section~\ref{sec:gammat_center}) ---  we follow \citetalias{y3-shapecatalog} in using a star weight for each PSF star. \citetalias{y3-shapecatalog} found that PSF stars are preferentially detected in regions with fewer galaxies due to the star-finding algorithm failing in crowded regions of the sky. Thus the number counts of the detected PSF stars is anti-correlated with the large-scale structure. This is unideal for the tangential shear test as we will then find a slight, negative correlation in the tangential shear around stars. Following \citetalias{y3-shapecatalog}, we correct for this effect by weighting all PSF stars with $w = n_{\rm gal}/n_{\rm star}$. We make a \textsc{Healpix} map of the galaxy number density, $n_{\rm gal}$, and of the star number density, $n_{\rm star}$, with a resolution of $\texttt{NSIDE = 512}$ (we have verified the robustness of our results to different choices of \texttt{NSIDE}). The weight map is then generated as the ratio of the two maps and stars are assigned the weight of the pixel corresponding to their position on the sky. These weights also have the secondary effect of optimizing our null test statistics by preferentially upweighting measurements at the locations of galaxies.

\subsection{Mean PSF in focal plane}
\label{sec:psf_map}

Figure~\ref{fig:psf_focalplane} shows a visual example of the PSF modeling errors in the data. In particular, it shows the residuals in the PSF model across the full focal plane of DECam, averaged over all $riz$-band exposures used in this work. We see consistent patterns when compared to DES Y3 \citep[see e.g., Figure 9 in][]{Jarvis2020}, \ie large-scale fluctuations in $e_{1, 2}$ due to the polynomial order used in the PSF model interpolation, and CCD-scale fluctuations in the PSF size $T_{\rm psf}$ (``tree rings'') due to impurities in the CCD silicon \citep{Estrada:2010:Treerings, Plazas:2014:TreeRings, Plazas:2014:TreeRings2}. The similarity between DES and \decade in the structure of these residuals is expected since the dominant contribution to these residuals are the optics and instrument systematics, which are the same across both datasets. Differences due to atmospheric conditions, which \textit{can} vary between DES and our dataset, are suppressed as we average over many exposures. Figure~\ref{fig:psf_focalplane} also shows that there are 1.5 additional CCD chips being used in our work compared to DES Y3 (the middle chip in the top row, CCD2, and the upper channel in the right-most chip in the 6th row, CCD31). DES Y3 took a conservative approach and discarded these data for shear measurements as the CCD chip/channel was not functioning for a subset of the survey period. In this work, we include these data as they have no problems apart from the fact that we discard data from these CCDs during the time period in which they malfunctioned. Note that these data have gone through extensive quality control and visual inspection (see Section~\ref{sec:desdm}).

\subsection{Brighter-fatter effect}
\label{sec:bf}
The brighter-fatter effect \citep[BF,][]{Antilogus2014, Guyonnet2015, Gruen:2015:BF} refers to a nonlinear sensor effect in the CCD, where the charge carriers induced by incoming photons are deflected by the accumulated charge at bright spots. A consequence of this effect is that brighter stars appear to be larger in the CCD images, which presents a challenge to PSF modeling as the models are built using the light profile around stars. To circumvent this effect, as well as other effects associated with bright stars (such as saturation), we do not use very bright stars for PSF modeling  (see Appendix~\ref{sec:psfex}). Furthermore, the CCD images are corrected to account for the BF effect \citep{Gruen:2015:BF}. Therefore, in this test we examine any remaining BF effect (after image corrections) in the stars that are used for PSF modeling.

\begin{figure}
    \centering
   \includegraphics[width=\columnwidth]{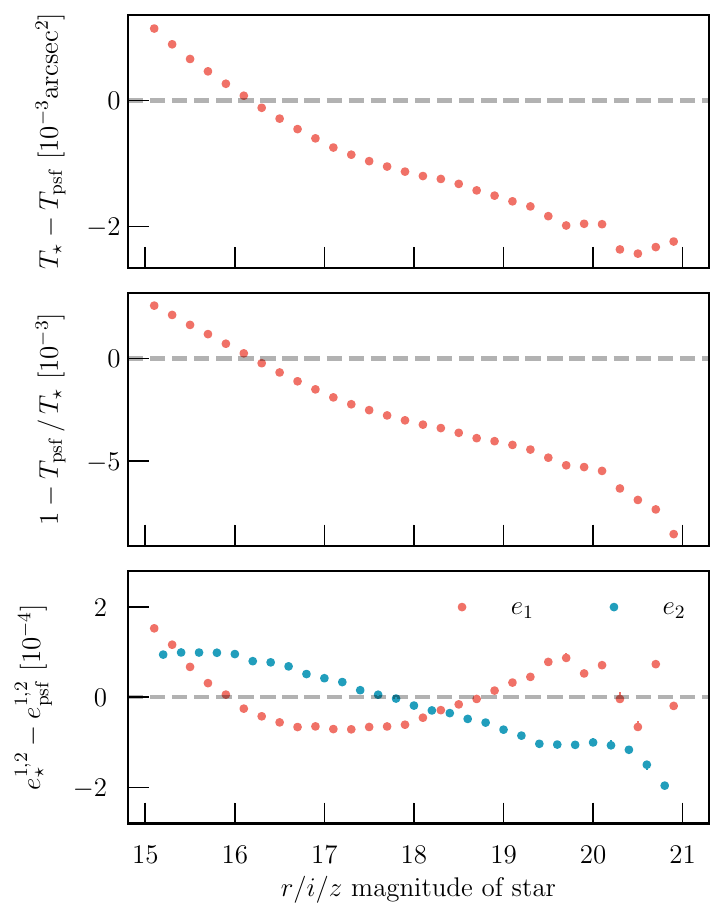}
    \caption{The brighter-fatter effect as found in the size residual (top panel), fractional size residual (middle panel), and shape residual (bottom panel), where the residuals are the difference between the star and the PSF model evaluated at the location of the star. }
    \label{fig:brighter_fatter}
\end{figure}

The BF effect is estimated by evaluating our PSF models at the location of the PSF stars and comparing the size and shape of the measured star to that inferred from the PSF model. The measurements of the star shape and size are performed via the \textsc{Galsim} package \citep{Rowe:2015:galsim}, using implementations of the HSM algorithms as developed in \citet{Hirata:2003:HSM, Mandelbaum:2005:HSM}. This comparison of the PSF model and the star measurement is done as a function of the star's magnitude in the relevant band. In Figure~\ref{fig:brighter_fatter}, we show the error and fractional error in the PSF model size (top and middle) and the error in the PSF ellipticity (bottom) as a function of the star brightness. The results combine measurements from the $riz$ bands, and the error bars --- which in almost all cases are much smaller than the points --- are derived via jackknife resampling from 150 spatial patches. The patch locations and assignments are constructed by passing the sky locations from the object catalog into the optimized k-means clustering algorithm in \textsc{TreeCorr} \citep{Jarvis2004TreeCorr}.

We observe a clear trend where the PSF size inferred from the model tends to be bigger (smaller) than the measured values from the fainter (brighter) stars. This is expected as the PSF model is constructed using stars of all magnitudes (\ie it is averaged across all magnitudes), and the BF effect will cause bright stars to appear larger, thus generating positive residuals towards the bright end in our plots. The fractional error in PSF sizes due to this effect is at the $10^{-3}-10^{-2}$ level, which would propagate as an additive bias to shear at the $10^{-5}-10^{-4}$ level (assuming the PSF ellipticity is at the level $10^{-2}$); see Equation~\eqref{eq:sys_breakdown}. The error in the $e_1$ and $e_2$ of the PSF model also shows a magnitude-dependent trend. These appear at the 10$^{-4}$ level, with $e_2$ monotonically decreasing with magnitude and $e_1$ having a more complicated shape. The amplitude in both cases is consistent to that found in DES Y3 \citepalias[][see their Figure 6]{y3-shapecatalog}. 

For both this test, and those to follow (particularly in Section~\ref{sec:psf_color} and \ref{sec:shearVsmcal}), we treat additive terms of $ \delta e \lesssim 2 \times 10^{-4}$ as negligible. This is motivated in three quantitative ways: (i) first, we expect a non-zero mean shear in our catalog due to cosmic variance. We estimate this for our data using jackknife resampling of the catalog, and find $\sigma(\langle \gamma \rangle) \sim 0.6\times10^{-4}$. This is broadly consistent with the value quoted in DES Y3, $\sigma(\langle \gamma \rangle) \sim 0.5\times10^{-4}$, which was estimated using simulated mock catalogs. Thus, additive terms with amplitudes at/below $\sim 2\times 10^{-4}$ are within the $3\sigma$ limit defined by cosmic variance. (ii) The mean shear of our catalog is of the order $\sim 5 \times 10^{-4}$ (this value is the average of $|\langle e_1\rangle|$ and $|\langle e_2\rangle|$, see Table~\ref{tab:neff_w2}). Following \citetalias{y3-shapecatalog}, the mean shear per-component is subtracted from the catalog before performing cosmic shear measurements. Thus, additive terms with amplitudes below this mean shear can be treated as negligible. Finally, (iii) the additive bias, $c$ (see Equation~\ref{eqn:mcal_bias}) estimated in the image simulations is of the same order, $\sim 5 \times 10^{-4}$ (Table~\ref{tab:ShearCalib}). Under these considerations, the trend of the mean PSF ellipticity with magnitude is negligible. Note that in Section~\ref{sec:rowe} we explicitly show the impact of these PSF model errors on the measurements relevant for cosmological inference, and find that the impact is negligible. Figure A1 in \citetalias{paper3} also shows that the impact of such model errors on our final cosmology constraints is negligible.

\subsection{PSF color dependence}
\label{sec:psf_color}

Next, we check the color-dependence of our PSF model. It is well-understood that the PSF is wavelength-dependent -- photons of different wavelength traverse differently through the atmosphere, optics, and detector, resulting in a different PSF at that specific wavelength. The PSF model is constructed using observations of stars, which generally have a more blue-tilted spectral energy distribution (SED) relative to galaxy SEDs. When applying this model for PSF-deconvolved shape measurements of galaxies, we are using a slightly incorrect model given the galaxy SED is different from the stars'. The effect is more severe when the filter band covers a broader wavelength range. A detailed investigation for the DES Y3 data can be found in \citet{Jarvis2020}, and an improved color-dependent PSF model is used in DES Y6 \citep{Schutt:2025:Y6PSF}. Our image data was already processed using a color-independent PSF model, but we quantify any residual PSF contamination in our shear estimates (\eg due to imperfect PSF modeling) using the Rowe/Tau statistics; see Section \ref{sec:rowe}. This is similar to the approach used in DES Y3, which also had a color-independent PSF model.

\begin{figure}
    \centering
   \includegraphics[width=\columnwidth]{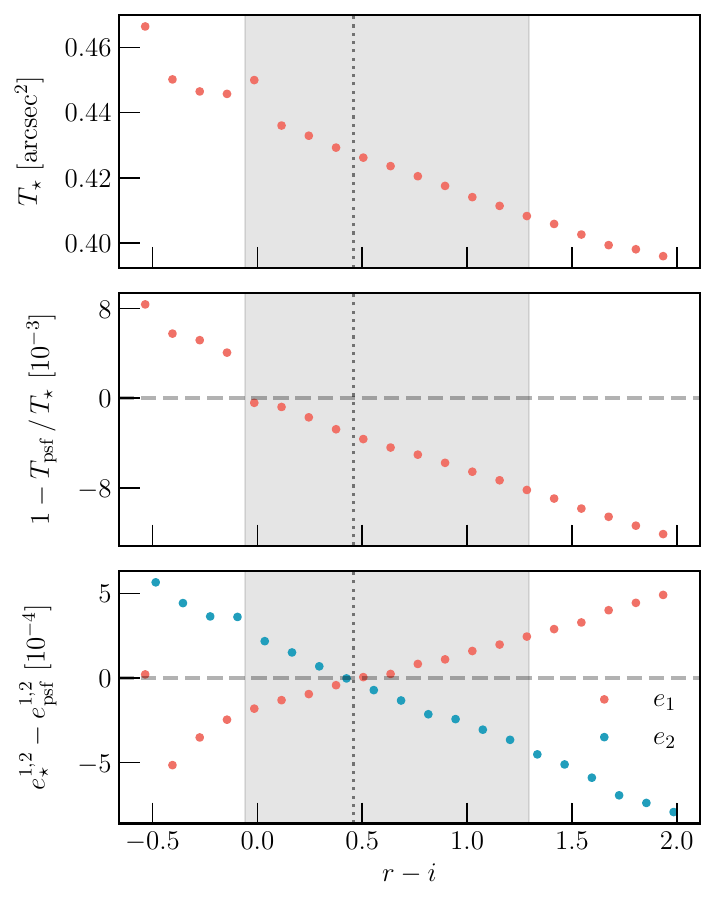}
    \caption{The variation in the PSF as a function of color, in this case $r-i$. The gray vertical band (dotted line) shows the 95\% region (median) of the $r-i$ color for the galaxy sample. The size varies by $0.02 \,\,{\rm arcsec}^2$ over most of the color range. The fractional error changes by $1\%$, and the shear error changes by around $2.5 \times 10^{-4}$.}
    \label{fig:psf_color}
\end{figure}

The most relevant contributor to the color-dependence is the effect of differential chromatic refraction \citep[DCR,][]{Plazas2012, Meyers2015}, where the PSF is preferentially elongated along the direction towards zenith with the elongation being stronger for sources with bluer colors compared to those with redder colors. This effect is particularly important when using the \decade shape catalog for cosmic shear as (i) the image data extends to higher airmass than DES and (ii) this effect could potentially imprint a spatially correlated bias instead of a stochastic, random bias.

Figure~\ref{fig:psf_color} shows the size of the PSF model and errors in the model size and ellipticity, all as a function of the $r-i$ color of the star. The gray band shows the 95\% region of colors spanned by the source galaxy sample. The PSF size error varies by $<1\%$ across this range, while the mean PSF ellipticity error has an amplitude below $2.5 \times 10^{-4}$ across this same range. Following the same argument in Section~\ref{sec:bf}, this residual error will not impact our cosmology analysis. In Section~\ref{sec:rowe} we quantify the impact of the additive PSF systematics on the data vectors relevant for cosmology. In \citetalias{paper3}, we also derive cosmology constraints using only the high/low airmass region of our survey and find the constraints are consistent with each other.

We note that the results of Figure~\ref{fig:psf_color} require a matching procedure. Since a PSF star is an exposure-level object, it only contains information about a single band. Thus, we have to assign a color to each star via a matching procedure on the sky. Briefly, we match stars across bands and exposures to assemble a ``catalog-level coadd'' and derive a color estimate averaged across exposures. We then assign that color estimate to each exposure-level star measurement. The catalog-level coadd is done by taking all stars in the (exposure-level) PSF star catalog and grouping together those stars with sky positions within a given \textsc{HealPix} pixel. We choose $\texttt{NSIDE} = 262144$ as it corresponds to a pixel scale of $0.8 \arcsec$, close to the $0.5 \arcsec$ matching radius used in other techniques within DES and our work. Each pixel --- which can now be treated as a separate star given the resolution of the map --- is then assigned a magnitude in each band by taking the weighted average of all available magnitude estimates of a given band, in that pixel. The weights used in the averaging are the signal-to-noise of each exposure-level measurement of the star. After assigning each pixel a set of $riz$ magnitudes (and the colors derived from the magnitudes), each star in the original, exposure-level catalog is assigned the colors of the pixel associated with its position on the sky.

\subsection{Mean shear as a function of \textsc{Metacalibration} quantities}
\label{sec:shearVsmcal}

\begin{figure*}
    \centering
    \includegraphics[width=2\columnwidth]{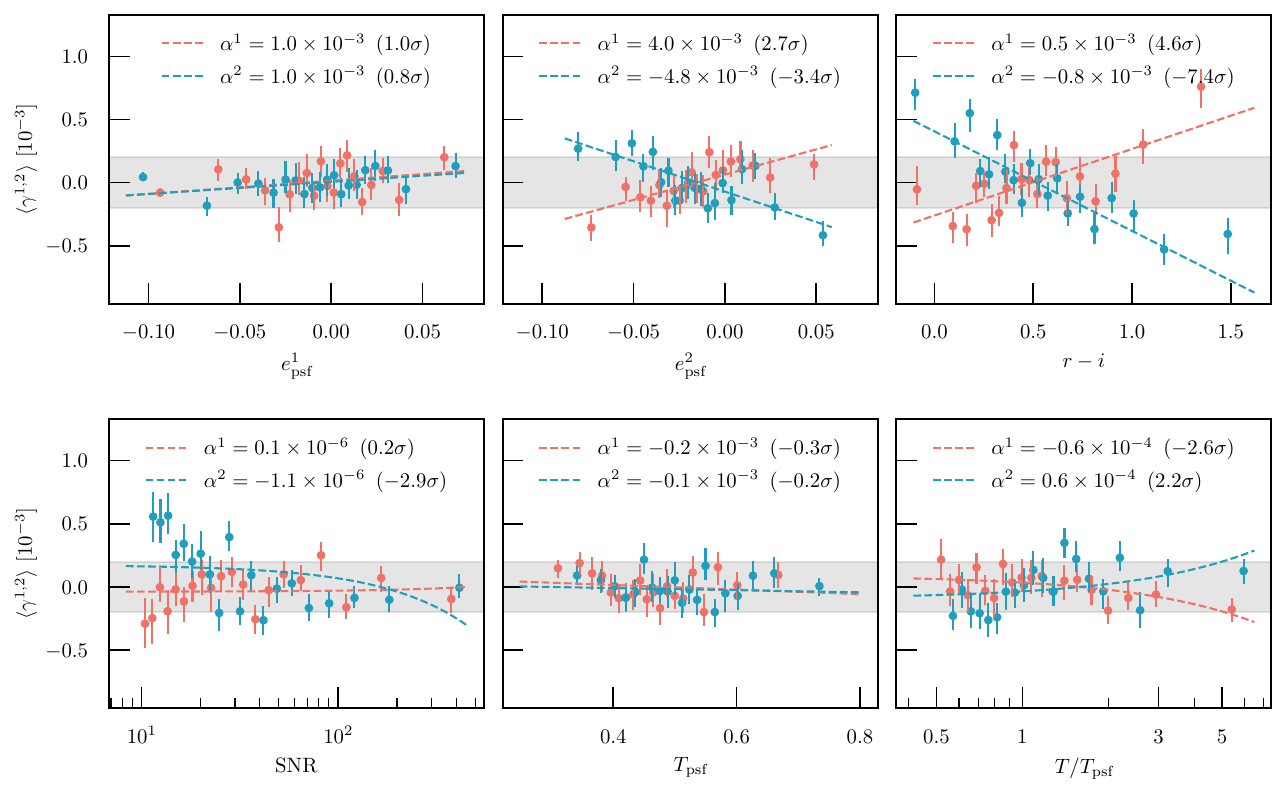}
    \caption{The mean shear of our fiducial sample, in percentile bins defined using other \textsc{Metacalibration} quantities. The mean shear is calibrated by a response that is computed bin-by-bin. Note that the PSF ellipticity and sizes used above are now measured within \textsc{Metacalibration}. The dots show the measurements while the dotted line is a linear regression. The legends quote the slope and the significance $\alpha/\sigma_\alpha$, where the uncertainty is estimated through 150 jackknife resamplings. The gray band shows the $\Delta e = 2\times10^{-4}$ region, and almost all points fall within this range.}
    \label{fig:shear_vs_X}
\end{figure*}

Next, we examine the correlation between our shear estimate with other quantities measured in \textsc{Metacalibration}, such as PSF size ($T_{\rm psf}$), PSF ellipticity ($e^{1, 2}_{\rm psf}$), galaxy color, galaxy signal-to-noise (SNR), and ratio of galaxy size to PSF size ($T/T_{\rm psf}$). We divide the galaxies into 20 equal-count (\ie percentile) bins, where the binning is done on the quantity we are correlating the shear estimate with. The mean shear in each bin is computed while using the shear weights and then calibrated using the response computed bin-by-bin. The response used here is the total response given in Equation~\eqref{eqn:combined_response} and also includes the selection term corresponding to the binning procedure. The correlation between shear and a given quantity is estimated via a linear regression. We fit the model
\begin{equation}
\langle \gamma^{1,2} \rangle = \alpha^{1,2} X + \beta^{1,2},
\end{equation}
where $X$ is the quantity of interest and $\alpha$, $\beta$ are the free parameters. A significant detection of $\alpha$ indicates that the mean shear is dependent on $X$. Figure~\ref{fig:shear_vs_X} presents the results of this test. The data points show the measured, mean shear per bin while the dashed line shows the fit from the regression. The slope of the linear fit and its associated significance are shown in the legend. The uncertainty used to quantify the significance is obtained via jackknife resampling with 150 spatial patches.

With few exceptions, the significance of the slope is consistent with zero within $3\sigma$. The most significant correlation is of shear with galaxy color, which is expected given the impact of DCR on the PSF modeling error and thereby the impact on the galaxy shapes. We note that the contribution of PSF correlations to the shear correlation functions is quantified, and found to be negligible, in Section~\ref{sec:rowe} below. Figure~\ref{fig:shear_vs_X} shows that the variation of the mean shear across bins for different quantities is generally within $\Delta e = 2 \times 10^{-4}$, which is negligible for the same arguments discussed in Section~\ref{sec:bf}. We note that there is a slight increase of the mean shear, $\langle \gamma_2 \rangle$, for objects at low SNR and this increase is similar to the result in DES Y3 \citepalias[][see their Figure 21]{y3-shapecatalog}, who confirmed that such an effect is completely negligible on cosmological constraints.

\subsection{Rowe statistics and Tau statistics}
\label{sec:rowe}

\begin{figure*}
    \centering
    \includegraphics[width=\linewidth]{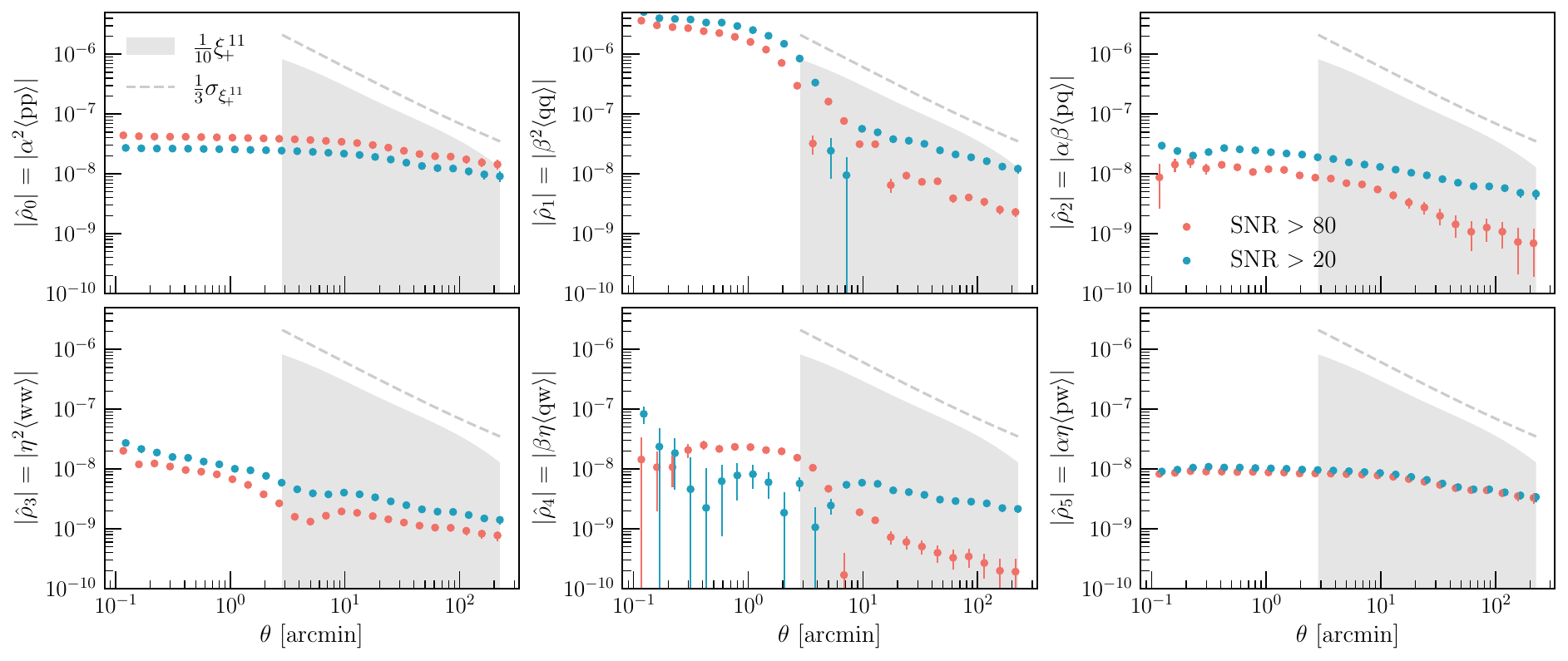}
    \caption{The $\xi_+$ component of the spin-2 correlation function defined by the PSF quantities: the ellipticity $p$, the ellipticity error $q$, and the size error $w$. We multiply the correlations by the coefficients $\alpha, \beta, \eta$ which quantify how the former propagate into the measured shear. See text for details on these coefficients. Here, we have used the 68\% upper bound (on the absolute value) for the coefficients rather than the best fit value as the former is a more conservative estimate. The gray band (dashed line) shows 10\% of the smallest cosmological signal (one-third of the smallest measurement uncertainty) in our data vector, from the autocorrelation of the 1st tomographic bin. In all cases the PSF signal is subdominant. This is true even if we vary the SNR threshold of the star sample used to estimate the Rowe statistics. This test serves an order-of-magnitude estimate of PSF contamination, and a test showing the negligible impact of this contamination on cosmology constraints is presented in \citetalias{paper3}.}
    \label{fig:Rowe_stats}
\end{figure*}

A key empirical test in all shear-related analyses is quantifying the impact of PSF systematic effects on the measured data vectors. PSF shapes can be correlated across the sky either due to the true PSF's intrinsic ellipticities (for example, from effects like DCR) or due to modeling/interpolation errors in the PSF model. The latter occurs because the PSF is modelled using test points (\ie stars) at a sparse set of locations on the focal plane and this model is then interpolated to the locations of galaxies. Thus, any errors in the interpolation will be shared by neighboring galaxies on the focal plane. The PSF modeling error is typically assumed to be additive, \ie our estimated shear $\bf{\gamma}$ differs from the true shear $\bf{\gamma^{\rm true}}$ by an additive factor that depends on the PSF model $\delta{\bf{e^{\rm sys}_{\rm psf}}}$ and noise. That is,
\begin{equation}
    \boldsymbol{\gamma} = \boldsymbol{\gamma}^{\rm true} + \delta \boldsymbol{e}^{\rm sys}_{\rm psf} + \delta \boldsymbol{e}^{\rm noise}.
\end{equation}
Motivated by \citet{Paulin-Henriksson2008} and \citet{Jarvis2016}, we parametrize $\delta \boldsymbol{e}^{\rm sys}_{\rm psf}$ via
\begin{align}
    \delta \boldsymbol{e}^{\rm sys}_{\rm psf} &= \alpha \boldsymbol{e}^{\rm psf} + \beta (\boldsymbol{e}^{\star} - \boldsymbol{e}^{\rm psf}) + \eta \left( \boldsymbol{e}^{\star} \frac{ \rm \it T^{\star} - T^{\rm psf}}{\rm \it T^{\star}}\right) \notag,  \\
    & \equiv \alpha \boldsymbol{p} + \beta \boldsymbol{q} + \eta \boldsymbol{w}.
    \label{eq:sys_breakdown}
\end{align}
That is, we identify three quantities that capture both the PSF model and its error, up to the second moment: $\boldsymbol{p} \equiv \boldsymbol{e}^{\rm psf}$ is the ellipticity of the PSF model; $\boldsymbol{q} \equiv \boldsymbol{e}^{\rm \star} - \boldsymbol{e}^{\rm psf}$ is the error in the ellipticity estimated as the difference between the ellipticity of the true PSF as measured from stars and of the PSF model evaluated at the location of the stars; and $\boldsymbol{w} \equiv \boldsymbol{e}^\star (\rm \it T^\star - T^{\rm psf} )/T^\star$ is the true PSF ellipticity multiplied by the fractional error in the size of the PSF model. Given these three quantities, we can write a set of six terms describing the two-point correlations between them \citep{Rowe2010, Jarvis2016}: 
\begin{align}\label{eq:rowe}
    \rho_0 & = \langle \boldsymbol{e}^{\rm psf} \boldsymbol{e}^{\rm psf}\rangle = \langle \boldsymbol{pp} \rangle,\nonumber\\
    \rho_1 & = \langle (\boldsymbol{e}^{\rm \star} - \boldsymbol{e}^{\rm psf}) (\boldsymbol{e}^{\rm \star} - \boldsymbol{e}^{\rm psf})\rangle = \langle \boldsymbol{qq} \rangle,\nonumber\\
    \rho_2 & = \langle \boldsymbol{e}^{\rm psf} (\boldsymbol{e}^{\rm \star} - \boldsymbol{e}^{\rm psf})\rangle = \langle \boldsymbol{pq} \rangle,\nonumber\\
    \rho_3 & = \langle (\boldsymbol{e}^\star (\rm \it T^\star - T^{\rm psf} )/T^\star)(\boldsymbol{e}^\star (\rm \it T^\star - T^{\rm psf})/T^\star) \rangle = \langle \boldsymbol{ww} \rangle,\nonumber\\
    \rho_4 & = \langle (\boldsymbol{e}^{\rm \star} - \boldsymbol{e}^{\rm psf}) \boldsymbol{e}^\star (\rm \it T^\star - T^{\rm psf})/T^\star \rangle = \langle \boldsymbol{qw} \rangle,\nonumber\\
    \rho_5 & = \langle \boldsymbol{e}^{\rm psf} \boldsymbol{e}^\star (\rm \it T^\star - T^{\rm psf})/T^\star \rangle = \langle \boldsymbol{pw} \rangle.\nonumber\\
\end{align}
These are commonly known as \textit{Rowe statistics} \citep{Rowe2010} and have been used in other works as diagnostics for the PSF modeling error \citep{Jarvis2016, y3-shapecatalog, Yamamoto2025, Schutt:2025:Y6PSF}. We note that recent analyses, motivated by the findings of \citet{Tianqing:2023:PSF}, have included additional terms in the Rowe Statistics that capture higher-order contributions \citep{Dalal2023, Li2023, Yamamoto2025, Schutt:2025:Y6PSF}. In this work, we follow the DES Y3 choices and do not include such higher-order terms. We note that \citet[][see their Figure 16]{Tianqing:2023:PSF} also show the impact of PSF contamination on cosmology is negligible regardless of whether these additional, higher-order PSF terms are included/excluded in the analysis.

The Rowe statistics measure the spatial correlations of the PSF model and of the modeling errors, summarized here via two-point statistics on the sky. These statistics, however, do not quantify the level of contamination this non-cosmological correlation introduces into the cosmic shear measurement, \ie it does not quantify the values of the coefficients $\alpha$, $\beta$, $\tau$ in Equation~\eqref{eq:sys_breakdown}. A second set of statistics, the \textit{Tau statistics}, is used to estimate these coefficients and thereby show the impact of PSF-related spatial correlations on the shear correlations. The Tau statistics are measured by correlating the three PSF-related quantities above with our galaxy shape catalog:
\begin{align}
\tau_0 &= \langle \boldsymbol{\gamma p} \rangle = \alpha \rho_0 + \beta \rho_2 + \eta \rho_5, \notag \\
\tau_{2} &= \langle \boldsymbol{\gamma q} \rangle = \alpha \rho_2 + \beta \rho_1 + \eta \rho_4, \notag \\
\tau_{5} &= \langle \boldsymbol{\gamma w} \rangle = \alpha \rho_5 + \beta \rho_4 + \eta \rho_3.
\label{eq:tau_stats}
\end{align}
The system of equations above can be solved to estimate $\alpha$, $\beta$ and $\eta$, and thereby the contamination of PSF-related spatial correlations in our shear correlation function.

\begin{figure*}
    \centering
    \includegraphics[width=\linewidth]{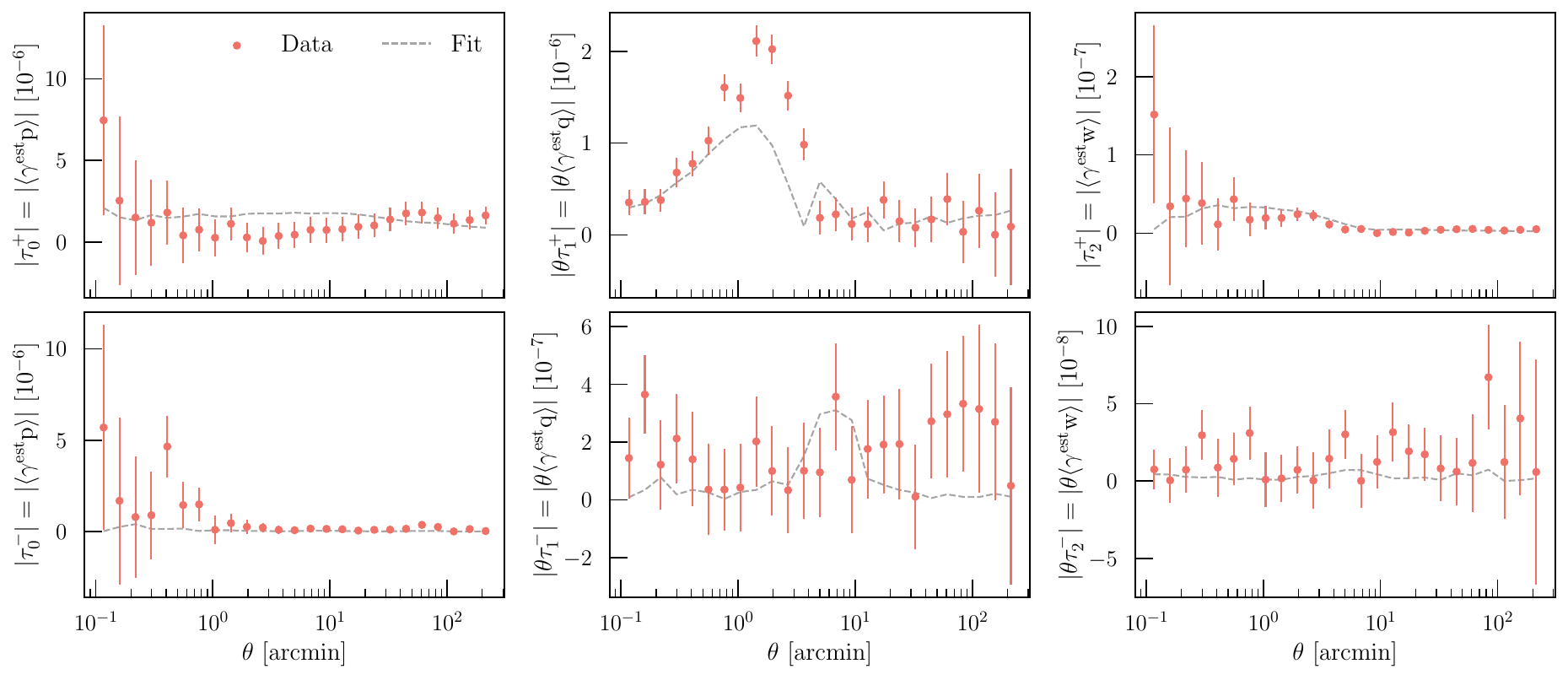}
    \caption{The Tau statistics of the shape catalog, which quantify the correlation between the estimated shear and the PSF terms. The red points show the measurement and the associated uncertainties, while the gray dashed lines are the fits to the measurement using the Rowe stats as templates; see Equation~\eqref{eq:tau_stats}. The fit is characterized by a p-value of $p = 0.46$. Note that the data points all have strong correlations with one another, as is also noted in \citetalias{y3-shapecatalog}.}
    \label{fig:Tau_stats}
\end{figure*}

The Rowe statistics of our catalog are estimated using \textsc{TreeCorr} \citep{Jarvis2004TreeCorr} and are presented in Figure~\ref{fig:Rowe_stats}. The error bars are generated via jackknife resampling of 150 patches.\footnote{We have checked that the results of the fits discussed below are similar even if we use $300$ or $450$ jackknife patches when estimating the covariance. The resulting values of $\alpha, \beta, \eta$ shift within their uncertainties, and we always find $p > 0.1$ for the best-fit. We account for the Hartlap factor \citep{Hartlap2007} in all cases. This indicates that any errors in our jackknife-based covariance estimate do not significantly affect the results of the test. We also note that recently, \citet{Guerrini:2024:UnionsRoweStats} introduced a semi-analytic method of covariance estimation that circumvents the limitations of the jackknife-based one. We have not used this technique as it was published after our catalog tests were already concluded, and also after our final cosmology results, in \citetalias{paper4}, were unblinded.} All quantities ($\boldsymbol{p}$, $\boldsymbol{q}$, $\boldsymbol{w}$ and $\boldsymbol{\gamma}$) are mean-subtracted before calculating the correlation and $\gamma$ is also calibrated using the response, $\langle \boldsymbol{R} \rangle$. The presented measurements in Figure~\ref{fig:Rowe_stats} are scaled by the relevant factors of $\alpha$, $\beta$ and $\eta$, and thus the lines represent the level of contamination introduced by the PSF-related spatial correlations. The overlaid gray bands show 10\% of the expected cosmic shear signal in the lowest redshift bin (\ie the bin where the amplitude is the lowest), using the redshift distribution found in \citetalias{paper2}. The gray dashed line also shows $1/3$rd of the expected uncertainty ($0.3\sigma$) on the shear measurement, estimated using the covariance model of \citetalias{paper3}. Both references are plotted only for the scales where we perform cosmic shear measurements ($2.5\arcmin < \theta < 250\arcmin$), but the Rowe statistics measurements extend to smaller scales. We observe that all the Rowe statistics are within the grey band, indicating the PSF model errors are subdominant to the cosmological signal. An explicit test of the impact of PSF contamination on cosmology constraints is shown in \citetalias{paper3}. We also tested for several SNR cuts in the star samples (see also Appendix~\ref{sec:psfex}) and find that the results from the Rowe statistics are relatively stable (\ie the correlations' amplitudes vary by less than a factor of 2) once we cut above SNR $>40$. Note that we use the star weights described above to weight the PSF stars, but our findings of negligible PSF contamination do not change if we ignore these weights.

Figure~\ref{fig:Rowe_stats} also shows that $\hat{\rho}_1$, which quantifies the autocorrelation of the PSF error, rapidly increases below $\approx 3 \arcmin$. This is expected and arises due to interpolation errors in the PSF model. The scale of the change is set by (i) the angular scale of an individual CCD image, and (ii) the order of the PSF interpolation polynomial. The same behavior has been found in other analyses of DES data \citep{Jarvis2016, Jarvis2020, y3-shapecatalog, Schutt:2025:Y6PSF} and is also described in \citet[][see their Section 5.3]{y3-shapecatalog}.

In Figure~\ref{fig:Tau_stats} we present the full set of Tau statistics together with the best-fit of the model described in Equation~\ref{eq:tau_stats}. The fit is done using a Monte Carlo Markov chain (MCMC) sampling routine from the \textsc{Emcee} package \citep{Foreman:2013:emcee}. The best-fit values are $\alpha = 0.0023 \pm 0.0031$, $\beta = 1.081 \pm 0.074$ and $\eta = -0.394 \pm 0.926$, with the p-value of the fit given as $p = 0.46$. For a DES-like sample, we expect $\alpha \sim 0$, $\beta \sim 1$ and $\eta \sim 1$; for example, see Table 2 in \citetalias{y3-shapecatalog}. Our estimates are consistent with these expectations.  

\subsection{Tangential shear around stars and field centers}
\label{sec:gammat_center}

Next we measure the mean tangential shear around PSF stars and field centers (\ie the center of an exposure in celestial coordinates). We expect both quantities to be zero in the absence of statistically significant, imaging-related systematics in the shape catalog. 

The tangential shear for a single source galaxy $i$ with respect to some reference location $j$ is defined as 
\begin{equation} 
\label{eq:tangential} 
    \gamma^{ij}_{t} = -\gamma_1^{i}\cos(2\phi^{ij})-\gamma_2^{i}\sin(2\phi^{ij}),
\end{equation}
where $\phi^{ij}$ is the angle between the line connecting the source $i$ and reference point $j$, and the declination-axis of the sky coordinate system \citep{Bartelmann2001}. In our case, the reference points are either field centers or stars. For the full shape catalog, we calculate
\begin{equation}
\label{eq:rangential_cat} 
    \gamma_t (\theta) =\frac{\Sigma_{ij}w^{ij}_{SX}\gamma^{ij}_{t, SX}(\theta)}{\Sigma_{ij}w^{ij}_{SX}(\theta)} - \frac{\Sigma_{ik}w^{ik}_{SR}\gamma^{ik}_{t, SR}(\theta)}{\Sigma_{ik}w^{ik}_{SR}(\theta)} ,  
\end{equation}
where the first term sums over all sources $i$ and reference points $j$ ($SX$ refers to source-reference point), and the second term sums over all sources $i$ and randoms $k$ ($SR$ refers to source-random). The source-galaxy shears are weighted by the \textsc{Metacalibration} weights $w^{i}_{S}$ (Equation~\ref{eq:weights}), are mean-subtracted, and are calibrated with the response. The reference points are weighted by $w^{j}_{X}$, and the specific choice of weights is described further below. The random points we generate are distributed uniformly in the shape catalog footprint and are weighted by $w^{j}_{R}$, which is proportional to the number of galaxies 
in that location of the sky. The combined weights are $w^{ij}_{SX} = w^{i}_{S} w^{j}_{X}$ and $w^{ik}_{SR} = w^{i}_{S} w^{k}_{R}$.   

\begin{figure}
    \centering
    \includegraphics[width=0.45\textwidth]{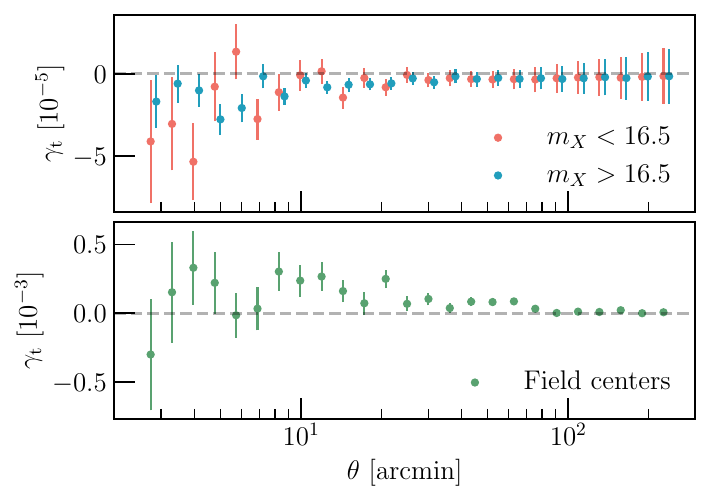}
    \caption{The tangential shear around PSF stars (top) and field centers (bottom). For the former, we measure the shear around both bright (red) and faint (blue) stars, split at $m_X \lessgtr 16.5$ in the relevant band. Field centers are the center (in celestial coordinates) of all exposures used in generating the shape catalog. The tangential shear around both stars and field centers is consistent with no signal, at p-values of 0.52, 0.11, and 0.24 for the bright star, faint star, and field center catalogs, respectively. Note that the datapoints are highly correlated, particularly on large-scales ($\theta \gtrsim 20 \arcmin$)}
    \label{fig:tangential_shear}
\end{figure}

In Figure~\ref{fig:tangential_shear} we show the tangential shear measurement for stars (upper panel) and the field centers (lower panel). All measurements are performed with \textsc{TreeCorr} \citep{Jarvis2004TreeCorr} and the error bars are estimated via jackknife resampling of 150 jackknife patches. Note that the datapoints on large-scales ($\theta > 20 \arcmin$) are all highly correlated.

For the PSF stars, we further split the star sample into a bright and a faint sample based on their magnitudes in the band corresponding to a given exposure. Following DES Y3 \citepalias{y3-shapecatalog}, we split on $m_X \lessgtr 16.5$ in band $X$. A non-zero tangential shear signal can be generated around bright stars due to significant artifacts (background subtraction, masking) near such stars that subsequently impacts the shear measurements of nearby galaxies. A tangential shear signal can also be generated around faint stars due to the standard PSF modeling errors we have discussed above. While the Rowe stats in Section~\ref{sec:rowe} have already validated our catalog, this test provides further validation as is it a different type of shear measurement. For both subsamples, we find no statistically significant tangential shear measurement. The $p$-values of the data vector are $p = 0.52$ and $p = 0.11$ for the bright and faint sample, respectively. 

In this test the PSF stars are weighted by $w_{X} = n_{\rm gal}/n_{\rm star}$ as discussed earlier. The random sample 
is weighted by $w_{R} = n_{\rm gal}$. The PSF star weights can be viewed as performing two distinct functions: (i) they transform a star sample that is spatially inhomogeneous into one that is uniformly distributed, and (ii) they upweight the sample in locations of galaxies so that the subsequent measurements are upweighted in those locations. The random sample we generate is already uniform and therefore does not need step (i). Thus, we perform only step (ii) for the randoms. 

As for the shear around field centers, a non-zero signal can indicate correlations of the measured shear with focal plane coordinates. For example, in Figure~\ref{fig:psf_focalplane} we have shown the correlation of PSF modeling errors with focal plane coordinates. However, there can exist other correlations as well. The field center catalog is the central R.A.\ and Dec.\ of the 42,700 exposures that correspond to the shape catalog described in Section~\ref{sec:shear}. Note that we remove all field centers that do not pass our foreground mask. The tangential shear measurement around field centers is consistent with zero with a p-value of $p = 0.24$.

Similar to the PSF star case, we generate weights for the field centers as $w_{X} = n_{\rm gal}/n_{\rm fc}$. Similar to $n_{\rm gal}$ and $n_{\rm star}$, $n_{\rm fc}$ is estimated as the number of field centers in a \textsc{Healpix} pixel of $\texttt{NSIDE = 512}$. All field centers are assigned an $n_{\rm fc}$ value from the pixel corresponding to their location on the sky. This term, $1/n_{\rm fc}$, is required in the weights as the locations of exposures used in our data are not uniform; given our use of data from community programs, which often target specific sources/regions, there are more exposures in regions with large-scale structure. Therefore the tangential shear signal around field centers will be contaminated by the actual cosmological signal. We correct for this using these field center weights. The randoms once again are weighted by $w_{R} = n_{\rm gal}$.

\subsection{Impact of observing conditions of the survey}
\label{sec:SysMaps}

\begin{figure}
    \centering
    \includegraphics[width = \columnwidth]{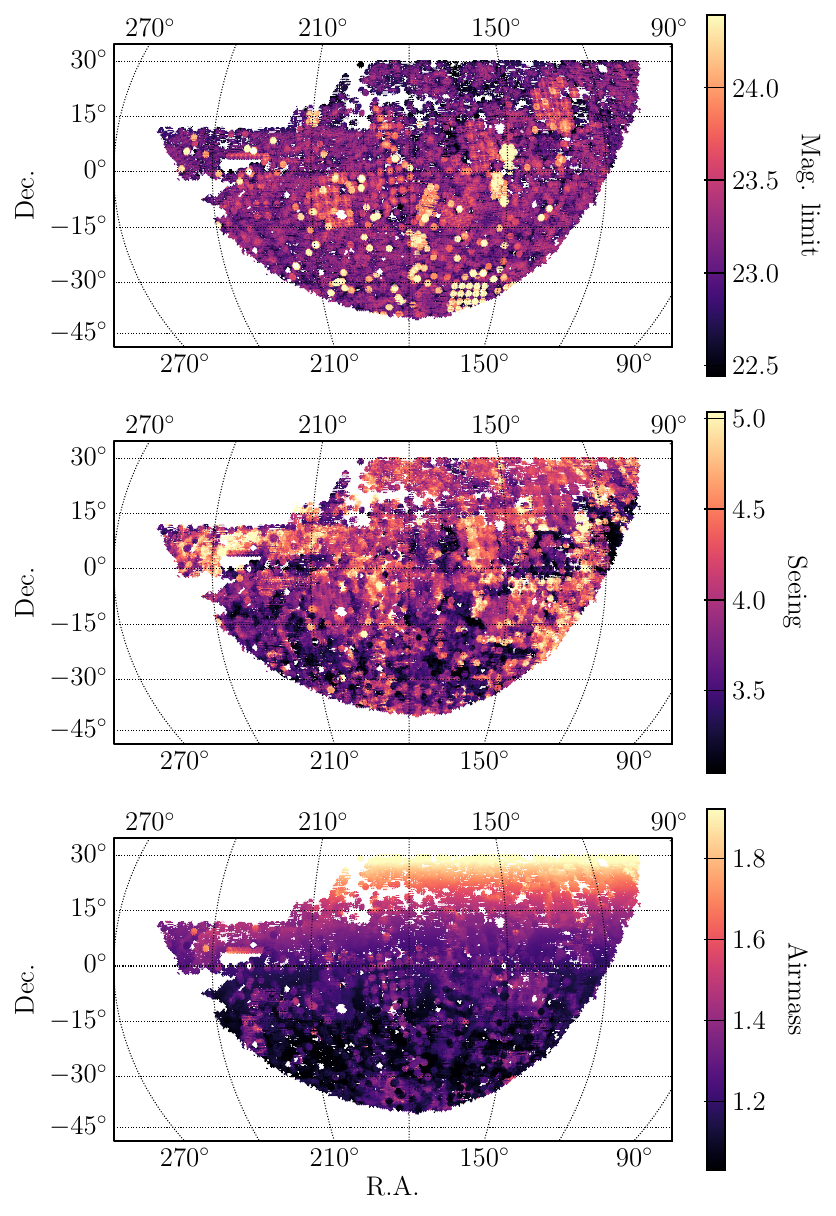}
    \caption{Example maps of the survey observing conditions. From top to bottom we show the magnitude limit (depth), the seeing (FWHM of PSF in pixel units), and airmass. All maps are shown in the $i$-band. Note that the source galaxy number density is significantly more homogeneous than the depth map (see Section~\ref{sec:diff2DES} and also Figure 1 in \citetalias{paper3}).}
    \label{fig:SP_maps}
\end{figure}

\begin{figure}
    \centering
    \includegraphics[width = \columnwidth]{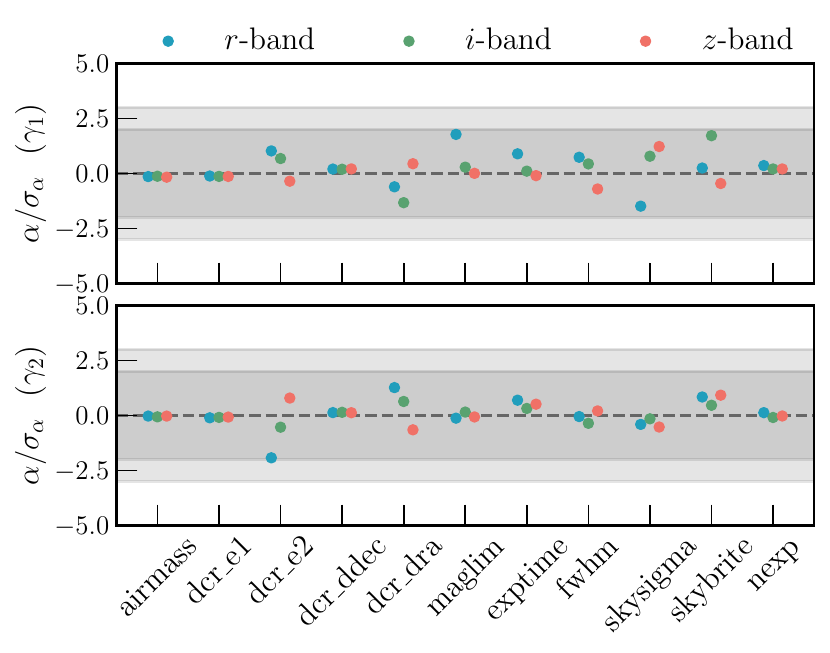}
    \caption{Significance of the slope ($\alpha$) of a linear fit between the mean shear and survey observing conditions. From left to right the observing conditions are airmass, DCR in the e1 and e2 directions (with response to focal plane orientation), DCR in the R.A. and Dec. directions, the magnitude limit (or depth), the exposure time, the seeing (PSF FWHM), the standard deviation of the sky background, the sky brightness, and the number of exposures. The top (bottom) panel shows correlations with $\gamma_1$ ($\gamma_2$). All slopes are consistent with zero within $2\sigma$ (dark gray band). The lighter band is the $3\sigma$ region.}
    \label{fig:Shear_vs_sys}
\end{figure}

We then test for any significant correlations between the estimated galaxy shear and various quantities related to the observing conditions of the \decade image data. Such a correlation could be non-zero if, for example, the PSF modeling error --- which is correlated with the estimated shear; see Equation~\eqref{eq:sys_breakdown} --- is correlated with an observing condition, such as the seeing or airmass. The PSF contamination effects, however, have been tested extensively in the sections above. Our test below probes the correlation between the shear measurements and a number of general observing conditions in our survey.   

We consider a range of quantities: airmass, DCR, magnitude limit (or depth), exposure time, seeing, sky variation, sky background, and exposure number counts, with each quantity also split across the individual $riz$ bands. 
All maps have a native resolution of $\texttt{NSIDE} = 4096$ and are generated from the individual CCD properties using the \textsc{Decasu} codebase\footnote{\url{https://github.com/erykoff/decasu}} as is done for DES Y6 \citep{Bechtol:2025:Y6GOLD}. Figure~\ref{fig:SP_maps} shows an example of some of these survey property maps.

To test the correlation of these maps with shear, we first make weighted-average maps of the response-calibrated $\gamma_{1}$ and $\gamma_{2}$ from the shape catalog. The response is computed pixel-by-pixel. To reduce the impact of noise in the response estimate, we take the average response in \textsc{Healpix} pixels of $\texttt{NSIDE} = 128$ (corresponding to $\approx30 \arcmin$). At this resolution, a single pixel contains roughly 4 CCD images. Our results below are qualitatively similar if we use $\texttt{NSIDE} = 256$. Given the maps of the survey properties and the mean shear, we fit a linear model between the pixel values of the two and extract the slope. We quantify the uncertainty on the slope using 150 jackknife patches. Figure~\ref{fig:Shear_vs_sys} shows the results for the detection significance of the slope, in all three bands. All slopes are consistent with zero within $2\sigma$. The largest significance is for the correlations with the DCR maps and for the bluest band (\ie the $r$-band correlation is larger than $i$-band and $z$-band). Note that the first 5 maps presented --- airmass, dcr\_e1, dcr\_e2, dcr\_ddec, dcr\_dra --- are all highly correlated as they derive from either the zenith angle or from both the zenith and the parallactic angles \citep[see Equations 7.6 and 7.7 of][]{Jarvis2020} of the observations. Overall, we find there are no statistically significant correlations of the mean shear with the survey observing conditions.

\subsection{$B$-modes}
\label{sec:b-modes}

\begin{figure}
    \centering
    \includegraphics[width = \columnwidth]{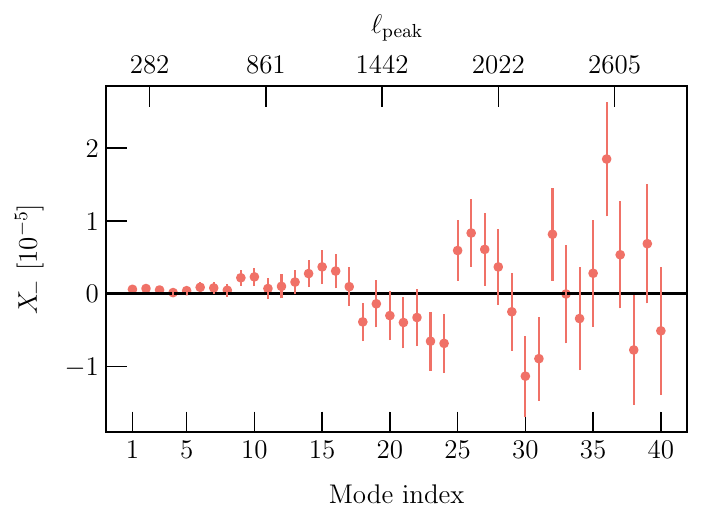}
    \caption{The $B$-mode amplitudes for 40 modes, estimated as a weighted sum of the $\xi_{+, -}$ correlation functions measured in 1000 bins between $2.5 \arcmin < \theta < 250 \arcmin$. The uncertainty estimates are obtained via 500 jackknife resamples. The measurements are consistent with null signal at $p = 0.11$, indicating this dataset has no statistically significant $B$-mode signal. The top axis shows the harmonic multipole corresponding to a given mode.}
    \label{fig:Bmodes}
\end{figure}

Finally, we measure the $B$-modes signal of our catalog, which is a test of other imaging systematics in the data. Shear is a spin-2 field and can be mathematically decomposed into a curl-free and a divergence-free component, which we commonly refer to as $E$-modes and $B$-modes, respectively \citep{Schneider2010}. To first order, gravitational lensing does not generate any $B$-modes,\footnote{At first order, the lensing field is the gradient of a scalar potential, and such a field has no curl mode by definition \citep{Schneider2010}.} and thus measuring a null signal in the $B$-modes is a powerful test of residual systematics in the shape catalogs. 

There are many different estimators for $B$-modes that can be used to test for systematic effects in a shape catalog \citep{Schneider2010, Becker2016, Alonso2019}. We use the \textsc{HybridEB} estimators developed in \citet{Becker2016}, which derive $B$-mode amplitudes from measurements of the real-space correlation functions. This estimator has the advantage of determining the minimum/maximum angular scale in real-space, allowing us to measure $B$-modes for only scales corresponding to the data vector used in the cosmology analysis (here we use $2.5 \arcmin  < \theta < 250 \arcmin$), and also allowing us to accurately incorporate the survey geometry without mixing $E/B$-modes. The $B$-mode amplitudes are computed as a weighted sum of the $\xi_{\pm}$ correlations. We measure these correlations in 1000 logarithmic bins between $2.5 \arcmin  < \theta < 250 \arcmin$ and then convert them into the $B$-modes amplitudes for 40 modes. The covariance matrix of $\xi_{\pm}$ is estimated via jackknife resampling and then transformed into that of the $B$-modes. This transformation is trivial as the $B$-mode amplitudes are a linear combination of the correlation functions. In \citetalias{paper3} we also estimate the $B$-modes for different tomographic bins of the shape catalog, using the harmonic-space estimator of \citet{Alonso2019}, and find no evidence for $B$-modes.

The jackknife covariance matrix of $\xi_{\pm}$ is now computed using 500 patches, rather than the fiducial choice of 150 patches. We found that using fewer realizations caused spurious, noisy correlations in the off-diagonal elements of the covariance that could then deteriorate the final $\chi^2$ of the null test. Given our footprint is $5,\!412 \deg^2$, each of the 500 patches is still sufficiently large with an area of $11 \deg^2$. We have also checked that the jackknife covariance matrix is qualitatively similar to those generated from simulation-based methods.\footnote{For these methods, we make 150 mock catalogs by (i) randomly rotating all the galaxies in our catalog to get the shape noise, and then (ii) assigning each galaxy a cosmological shear based on its position, using lensing maps generated from the \textsc{CosmoGrid} simulations of \citet{Kacprzak2023Cosmogrid} or the \textsc{Ulagam} simulations of \citet{Anbajagane2023Inflation}; the former has a slightly higher resolution while the latter has a factor of $\mathcal{O}(20)$ more independent sky realizations at fixed cosmology. The covariance is estimated by measuring $B$-modes for each of the mock catalogs.} We use the jackknife estimate of covariance for our null-tests, given this estimate is more reliable for capturing any/all effects in the dataset that contribute to the final covariance.

Figure~\ref{fig:Bmodes} shows the $B$-mode amplitudes, denoted as $X_{-}$ following the nomenclature of \citet{Becker2016}. The measurement is consistent with the null signal at a p-value of $p = 0.11$. Therefore, we do not find $B$-modes in our catalog. We have also checked that the contribution of the PSF two-point correlation functions (Equation~\ref{eq:rowe}) to the $B$-modes is at least one to two orders of magnitude below the measurement above. As mentioned above, we also compute the $B$-modes per tomographic bin in \citetalias{paper3} and find the measurement is consistent with no detection for all auto- or cross-correlation measurements.

\section{Shear calibration bias}
\label{sec:calibration_bias}

While in Section~\ref{sec:sys_tests} we directly use the shape catalog to check for systematic biases, in this section we validate and calibrate our shear estimator. This is done by testing our shear measurement algorithms on image simulations, where the true shear of the sample is known. Similar to the approach in DES Y3 \citep{Maccrann2022ImSim}, these tests first validate the performance of the \textsc{Metacalibration} algorithm in ideal, simplistic simulations and then quantify the final uncertainties/biases on shear calibration using realistic image simulations.

The \textsc{Metacalibration} estimate described in Section~\ref{sec:metacal} can be biased. Generically, this bias is written as,
\begin{equation}\label{eqn:mcal_bias}
    \gamma_i^{\rm est} = (1 + m)\gamma_i^{\rm true} + c
\end{equation}
where $m$ is the multiplicative bias, and $c$ is the additive bias \citep{Heymans:2006:Bias}. In ideal conditions --- with minimal masking, no blending, no detection/selection biases, \textit{etc.} --- the \textsc{Metacalibration} estimator is biased at $m \sim \mathcal{O}(10^{-4})$, which a negligible effect \citep{Sheldon2017}. However, under realistic conditions where sources are partially masked, blended etc., the multiplicative bias can be at the percent level. Estimates of additive bias in simulations are less vital as we subtract by the mean ellipticity of the catalog from each shape estimate, and thus the impact of the additive bias will be significantly reduced. Therefore, the calibration efforts in the community generally focus on the multiplicative bias, and we take the same approach here.

We use a suite of image simulations --- that we specifically develop to mimic the processing framework of the DESDM pipeline used in the \decade project --- to calibrate these multiplicative biases. We describe the simulation pipeline in Section~\ref{sec:ImsimPipeline} and validate it in Section~\ref{sec:ImsimValidation}. Finally, in Section~\ref{sec:ImsimResult}, we present the main result of our calibration: the final estimates for residual shear calibration bias, $m$, and their uncertainties, $\sigma(m)$. These will be an input to the final cosmology analysis \citepalias{paper4}.

\subsection{Image simulations pipeline}\label{sec:ImsimPipeline}

Our pipeline is developed following the approach of \citet{Maccrann2022ImSim}. Briefly, the pipeline fully simulates CCD images by injecting synthetic galaxies and star models, adding noise, masks, background etc. associated with the image. Then, the entire DESDM pipeline is run on these CCD images to reproduce coadded images, \textsc{SourceExtractor} detection catalogs, \textsc{Metacalibration} shear estimates etc. Other surveys have used similar approaches, though with differences in the exact, technical details; see \citet{Li2022} and \citet{Li:2023:Imsims} for the pipelines of HSC and KiDS, respectively. We now describe each piece in more detail:

\textbf{Coadd tile selection:} We randomly select 1600 coadd tiles from across the footprint. A subset of the galaxies in these tiles are in regions removed according to our foreground mask, and thus are not used in the calibration.

\textbf{Galaxy sample:} We use the same galaxy sample as \citet{Maccrann2022ImSim}, which was obtained from one of the DES deep fields that overlapped with the COSMOS field \citep{Scoville:2007:COSMOS} and was imaged using the Hubbe Space Telescope (HST). The availability of HST images with their significantly smaller PSF allows for the precise measurement of the galaxy shapes/morphology, while the availability of the DES images allows for measurements of DES photometry for these same galaxies. See \citet{Hartley:2022:Y3Deepfields} for more details on the DES-COSMOS deep fields catalog used as our input galaxy sample. \citet{Maccrann2022ImSim} perform additional quality cuts to obtain a purer sample of galaxies, and the same has been done for this work. In practice, we select all galaxies brighter than $i < 25.5$, which is two magnitudes deeper than the median survey depth of $i \approx 23.5$ (see Figure~\ref{fig:mag_dist_cuts}) and two magnitudes deeper than our $i$-band selection cut (see Section \ref{sec:selection}). The galaxy models are obtained from the \textsc{Fitvd} method \citep{Bechtol:2025:Y6GOLD}, which fits an exponential profile and deVaucouleurs profile to each galaxy image, representing the galactic disk and bulge, respectively. We use the \textsc{Ngmix} \citep{Sheldon:2015:ngmix} and \textsc{Galsim} \citep{Rowe:2015:galsim} software packages to ingest the galaxy morphology and render an image of the galaxy.

\textbf{Galaxy counts, shapes \& positions:} The number of galaxies in each tile is drawn from a Poisson distribution with a mean of $\langle N_{\rm gal}\rangle = 160,\!000$. After processing the images through the DESDM pipeline and performing all cuts shown in Figure~\ref{fig:mag_dist_cuts}, we obtain a sample with $n_{\rm gal} \approx 5\,\, {\rm gal}/{\rm arcmin}^2$, similar to the raw number density of our data as shown in Table~\ref{tab:neff_w2}. First, all galaxy shapes are randomly rotated. Then, we apply a known, constant shear to all rotated galaxies. We chose $\gamma_1 = \pm 0.02$ and $\gamma_2 = 0$, following \citet{Maccrann2022ImSim}, and the two variants ($\pm 0.02$) will be used for a noise suppression step discussed further below. The galaxy positions are placed randomly on the tile, and thus galaxies are allowed to be blended. Note that in the real sky the galaxy positions are correlated, which results in less blending of objects at similar redshifts. This was not incorporated in \citet{Maccrann2022ImSim} and is also not included in this work. However, other approaches include cosmologically consistent spatial correlations in their image simulations by sampling sky positions from halo catalogs in N-body simulations \citep[\eg][]{Euclid:2019:Imsims, Li:2023:Imsims}. In Section \ref{sec:calibration_bias}, we discuss the consequences from not including such spatial clustering, and how our analysis accounts for any biases from this choice.

\textbf{Simulated star catalog:} To improve the realism of our synthetic images, we inject simulated stars. We use the star catalog of \citet{DalTio:2022:StarSimLSST}, who developed this dataset for LSST using a modified run of \textsc{Trilegal} \citep{Girardi:2012:Trilegal}. The catalog photometry is provided in the LSST bands and these were converted to the DES bands, similar to what is done in \citet{Yamamoto2025}. For each simulated coadd tile, we select all stars whose positions in the \textsc{Trilegal} catalog are within the tile boundaries. However, we then randomly draw new positions (from a uniform distribution in coadd pixel coordinates) for each star and use this as the location of the injection. For binary star systems in the catalog, we also randomly sample the positional angle of the system on the sky. Similar to the galaxy injections, we subselect stars with $i < 25.5$ to form our injection sample.

\textbf{Extinction:} All objects are reddened during injection. The extinction coefficients, $E(B - V)$ are taken from \citet{Schlegel:1998:Dust}. The reddening coefficients, $A_b$, for band $b$, are 2.140, 1.569, 1.196 for the $riz$ bands, respectively \citep[][see their Section 4.2]{DES2018}. We obtain the reddened magnitudes, $m_{b} = m_{b, 0} + A_b \times E(B-V)$, by computing $A_b \times E(B-V)$ at the sky location of each object we inject.

\textbf{Simulating individual CCD images:} Once we have defined the shapes and location of galaxies --- and also the locations of the stars --- on the coadd tile, we simulate the individual CCD images that contribute to the coadd image. The objects are rendered as \textsc{Galsim} models and then convolved with the \textsc{PSFEx} model of that CCD image, where the latter model is evaluated at the location of the injection. These PSF-convolved galaxy models are drawn onto an CCD image that is initially blank. We then add to it the background and noise associated with the original image. The noise is simulated using the image weights given the latter is an estimate of the noise variance. The simulated image is also assigned the same weights and mask as the original.

\textbf{Generating coadd image:} The individual images are passed through a pixel-correction step with DESDM, called \texttt{coadd\_nwgint}, and these processed images are then coadded using the \textsc{Swarp} \citep{Bertin:2010:Swarp} algorithm to form coadd images for each of the three bands ($r, i, z$). The images from the three bands are then coadded again, using \textsc{Swarp}, to form the detection coadd.

\textbf{\textsc{SourceExtractor} object catalogs and \textsc{Meds} files:} The detection coadd is analyzed using \textsc{SourceExtractor} \citep{Bertin:1996:SrcExt} to obtain the object catalog. The algorithm is then run three more times in ``dual mode'' to get the object properties (photometry, size etc.) in each of the three bands. This catalog, alongside the simulated images, is used to create a  multi-epoch data structure file \citep[MEDS,][]{Jarvis2016} for the coadd tile.

\textbf{\textsc{Metacalibration}}: Finally, we process the simulated \textsc{Meds} file through the same shape measurement pipeline as that used on the data (see Section ~\ref{sec:metacal}).

At the end of this process, each simulated coadd tile will have two catalogs associated with it. One from a simulation where all galaxies are sheared with $\gamma_1 = + 0.02$ and another with $\gamma_1 = -0.02$. Both simulations contain the same rotated galaxies, noise realisations etc. and differ only in the applied external shear. Together, the two runs are used to estimate the multiplicative bias $m$ via the variance cancellation technique of \citet{Pujol:2019:Imsims}. This cancellation is done by using both versions of the tile to estimate the shear of the sample,
\begin{equation}\label{eqn:pujol_shear}
    \gamma_i^{\rm est} = \frac{\langle e^+\rangle - \langle e^-\rangle}{\langle R \rangle},
\end{equation}
where $\langle e^{\pm}\rangle$ is the mean ellipticity in the simulation with $\gamma_{1\pm} = \pm 0.02$, and $\langle R \rangle = \langle R^+ \rangle + \langle R^- \rangle$. Since both simulated versions only differ in the external shear applied to galaxies, all noise factors can be suppressed by taking the difference between the two. In practice, we find the estimator in Equation~\eqref{eqn:pujol_shear} improves our precision by at least a factor of 2, and often more. Once $\gamma_i^{\rm est}$ is estimated, we can infer $m$ using Equation~\eqref{eqn:mcal_bias} with $\gamma_1^{\rm true} = 0.02$. We always use $\gamma_1$ ($\gamma_2$) for estimating the multiplicative (additive) bias. The uncertainties on $m$ and $c$ are obtained through jackknife resampling by dropping the objects from one coadd tile at a time.

Our image simulation pipeline is also modified to perform synthetic source injection --- analogous to the synthetic source injection pipeline in DES denoted as ``\textsc{Balrog}'' \citep{Suchyta:2016, Everett:2022, Anbajagane:2025:Y6Balrog} --- and is used for the photometric redshift calibration presented in \citetalias{paper2}. In that work, we run our pipeline on the real (not simulated) CCD images from DECam and show we reproduce the object fits from the DESDM pipeline, which is a validation our pipeline accurately mimics the one used on the real data.

\subsection{Validation}\label{sec:ImsimValidation}

The image simulations use an extensive, multi-faceted pipeline. In \citetalias{paper2}, we show that our pipeline matches the image processing done on the actual data. However, there are still additional subtleties involved with simulating synthetic sources. Past works have performed an end-to-end validation of their simulations by checking the pipeline produces small multiplicative biases, in specific, idealized cases. We show these results in Section~\ref{sec:ImsimTest}. Furthermore, we check the realism of our simulated dataset in Section~\ref{sec:ImsimVsData} by comparing the properties of simulated galaxies with those of the real galaxies in our data.

\subsubsection{Unit tests}\label{sec:ImsimTest}

Following \citet{Maccrann2022ImSim}, we perform a number of units tests to validate the simulation pipeline. This involves running the simulation pipeline in a series of simplistic settings. The galaxies in these tests are all placed on a square grid (\ie there is no blending of galaxies) and they have exponential profiles, with half-light radii that we either fix to $0.5\arcsec$ or vary by sampling radii from the \textsc{Cosmos} catalog. In the latter case, we always limit the maximum radii to $1\arcsec$ in order to prevent overlap in galaxy injections. All injected galaxies also have the same brightness in all three bands ($riz$), and we vary the galaxy magnitude in different runs. We test one run where the PSF is a Gaussian with FWHM of $0.9\arcsec$, while the rest of the runs use the fiducial \textsc{PSFEx} model used in the main simulations. In all validation runs, we do not perform \textsc{SourceExtractor} detection, as any such detection step can introduce a shear-dependent detection that will lead to residual multiplicative biases \citep[][see their Section 4]{Maccrann2022ImSim}. Instead, we run the pipeline in ``true detection'' mode where the exact injected locations of objects are passed back into the pipeline as a detection catalog. This is done \textit{only} for the unit tests in this subsection. The fiducial run, whose results are discussed in Section~\ref{sec:ImsimResult}, include the \textsc{SourceExtractor} step as part of the full DESDM pipeline.

Table~\ref{tab:Imsim_unittest} shows the results from our unit tests. We show only the multiplicative bias, $m$, as it is the key quantity to validate. The order-of-magnitude expectation for the minimum bias in the simulated lensing signal is $m \sim \gamma_1^2 \sim 10^{-4}$, which arises because our shear estimator ignores terms that are quadratic, or higher, in $\gamma_1$ \citep{Bernstein:2016:Shear, Sheldon2017}. Our unit tests show the multiplicative bias is below $m < 10^{-3}$ at $>3\sigma$ significance, which is the requirement for our validation, and some tests find no bias down to the $10^{-4}$ level. The additive bias, $c$, is not quoted above but is at/below $10^{-4}$ for all simulation versions, similar to the amplitudes seen in previous works \citep[][see their Table B1]{Maccrann2022ImSim}.

\begin{table}
    \centering
    \begin{tabular}{c|c|c|c|c|c}
        \hline
        No. & mag & size & PSF & Mask & $m \,(10^{-3},\,\, 3\sigma)$ \\
        \hline
        1 & 17 & $0.5\arcsec$ & $0.9 \arcsec$ (Gauss) & No & $0.146 \pm 0.486$ \\[2pt]
        2 & 17 & $0.5\arcsec$ & \textsc{PSFEx} & No & $-0.006 \pm 0.413$ \\[2pt]
        3 & 17 & variable & \textsc{PSFEx} & No & $0.040 \pm 0.326$ \\[2pt]
        4 & 20 & variable & \textsc{PSFEx} & No & $0.001 \pm 0.416$ \\[2pt]
        5 & 17 & variable & \textsc{PSFEx} & Yes & $-0.080 \pm 0.351$ \\[2pt]
        \hline
    \end{tabular}
    \caption{Results for the validation tests on the simulation pipeline that is then used to calibrate the shear measurements. All unit tests show $m < 10^{-3}$ with $>3\sigma$ confidence, indicating the pipeline is robust. See Section~\ref{sec:ImsimValidation} for details on the tests.}
    \label{tab:Imsim_unittest}
\end{table}

\subsubsection{Comparing image simulations with data}\label{sec:ImsimVsData}

\begin{figure*}
    \centering
    \includegraphics[width 
=1.5\columnwidth]{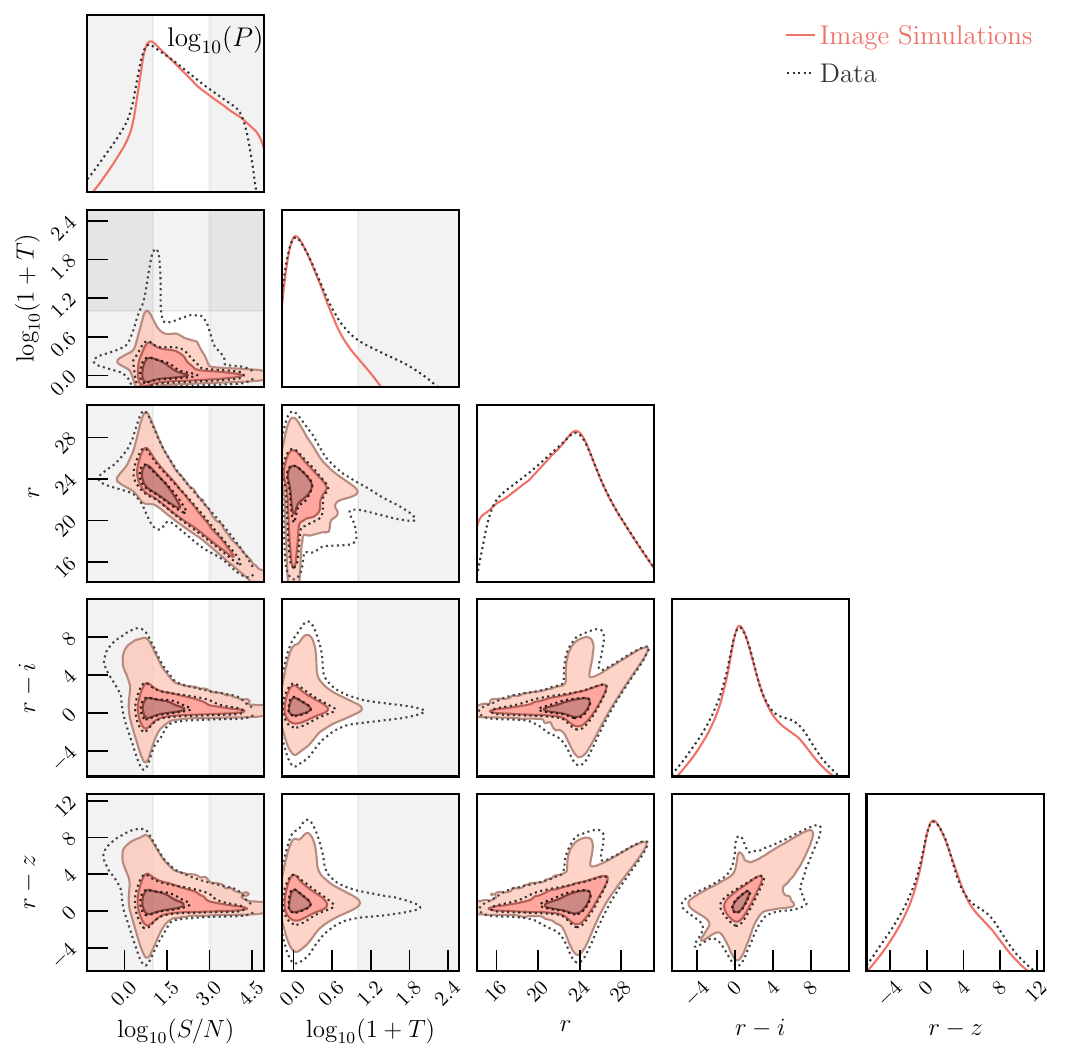}
    \caption{Distribution of \textsc{Metacalibration} properties of objects from the simulations (red) and data (grey). We use the raw catalogs and have not applied our selection function. The contours show the $1\sigma$, $2\sigma$, and $3\sigma$ regions. The 1D distributions along the diagonal are the log probability. The simulations are an excellent match to the data until the $2\sigma$ region, and generally still a good match up to $3\sigma$, with the largest deviations coming from the fact that the COSMOS catalog used for simulated injections lacks large (high $T$) objects. Note that this is the raw catalog with no cuts. The selections for signal-to-noise and size are shown in the light gray bands to guide the reader. We emphasize that the visually apparent disagreement is between the $3\sigma$ contours, while the $1\sigma$ and $2\sigma$ regions of the two distributions are nearly indistinguishable in most panels.}
    \label{fig:Imsim_Data_compare}
\end{figure*}

While Section~\ref{sec:ImsimTest} validated the simulation pipeline, an equally important validation is checking that the object catalog from the simulations is representative of the actual data. Figure~\ref{fig:Imsim_Data_compare} shows a corner plot comparing the distributions of the \textsc{Metacalibration} properties in the simulations and in the data. Both samples have no selections applied to them. The 2D distributions show the $1\sigma$, $2\sigma$, and $3\sigma$ contours. The 1D distributions along the diagonal show an excellent match between simulations and data; the sole exception is the size distributions where the simulations show a deficit of large objects (high $T$ values). This is a known difference, previously seen in \citet{Maccrann2022ImSim}, and arises from the rarity of large objects in the \textsc{Cosmos} catalog from which our injection sample is derived \citep{Hartley:2022:Y3Deepfields, Maccrann2022ImSim}. Note that the 1D distributions are presented on a \textit{logarithmic} scale, and thus the difference is amplified visually. On the other hand, the $68\%$ and $95\%$ contours of the distributions are on top of each other, showcasing an excellent match. Thus, while there is indeed a deficit of large objects in the simulated catalogs, the majority of the catalog is still representative of the data. Therefore, the simulations match the data and therefore can be relied on for calibrating the shear estimates. 

To be conservative about the consequences from the deficit of large objects, we also quantitatively check the deficit's impact on our estimates of $m$. We take all simulated galaxies (post selection cuts) and reweight them by an empirical function such that their size distribution now matches that of the data. We then confirm that the two variants of $m$ --- our modified, up-weighted estimate here, and the fiducial estimate presented in Table \ref{tab:ShearCalib} --- are consistent within $0.1\%$, or $0.3\sigma$ of the calibration uncertainties, for all four tomographic bins. The deficit of large objects (which populate only the tail of the sample distribution) have a negligible impact on our estimate of $m$. As we discuss below, our final cosmology constraints in \citetalias{paper4} are insensitive to changes of $\mathcal{O}(1\%)$ in $m$.

A salient point is also the realism of the PSF used in our image simulation. In our simulation pipeline, the \textsc{PSFEx} model is assumed to be the true PSF; we generate mock observations of a galaxy using the \textsc{PSFEx} model and analyze said observations using the exact same model. Any systematic bias in the former (and therefore any downstream effects of the bias on the final shear calibration) will not be captured in the simulations due to this assumption. We have already presented some signs of such an error, in Figure~\ref{fig:brighter_fatter} and Figure~\ref{fig:psf_color}. However, we now explicitly compute the average size error for our galaxy sample, by using the $1 - T_{\rm psf}/T_{\star}$ of our PSF star sample and calculating a weighted average across the catalog. The weights are the star weights discussed previously which upweight the signal at the location of galaxies. We find the size error is $-0.2\%$, \ie the PSF model is on average 0.2\% larger than the size of the stars.

We check the impact of this mean error by simulating it in our pipeline: we continue to convolve the galaxy with the \textsc{PSFEx} model before injection, but provide a slightly smaller PSF to \textsc{Metacalibration} during the shape fitting procedure. The recovered multiplicative bias is $m = (2.6 \pm 0.2) \times 10^{-3}$, which is within the uncertainties of our shear calibration as we will discuss below.

\subsection{Final shear calibration}\label{sec:ImsimResult}

For the final calibration of the data, we take the 1600 tiles of our fiducial simulation run and compute the multiplicative bias using only galaxies that pass the cuts discussed in Section~\ref{sec:shear}. We replicate all cuts described in that section, with the sole exception of the \textsc{Fitvd} failure cut which is left out as we do not run \textsc{Fitvd} fitting in our pipeline. This minor mismatch, of not using \textsc{Fitvd} flags, is similar to that done in the DES Y3 calibration work \citep{Maccrann2022ImSim}.

Furthermore, we require a calibration factor per tomographic redshift bin. These bins have been defined in \citetalias{paper2} by using self-organizing maps \citep[SOMs;][]{Kohonen:1982:SOM, Kohonen:2001:SOM}, which were trained on galaxy photometry, to assign galaxies to tomographic bins. We take the SOM built from the data and use it to assign simulated galaxies --- via their \textsc{Metacalibration} photometry estimates --- to the four tomographic bins. We then estimate the multiplicative bias in each bin using Equation~\eqref{eqn:mcal_bias} and Equation~\eqref{eqn:pujol_shear}. Note that when making these estimates, all simulated galaxies are assigned weights using the 2D grid defined in Section~\ref{sec:weights}, and the final calibration per bin is obtained via weighted averages of the relevant quantities.

Table~\ref{tab:ShearCalib} shows our final results for the calibration factors. Our estimates for $m$ have a similar amplitude to that of \citet[][see their Table 3]{Maccrann2022ImSim}. We note also that our estimate of $m$ in bin 3 has a higher amplitude than that of $m$ in bin 4, but the two estimates are only $0.5\sigma$ from each other so this ordering is not statistically significant. We also note that our estimate of the additive bias, $c$ --- which is extracted from the simulations using measurements of $\gamma_2$ --- has a somewhat similar amplitude as the mean shear of $\gamma_2$ measured in the data (see Table~\ref{tab:neff_w2}).

We note two caveats for these results. First, the DES Y3 image simulations campaign \citep{Maccrann2022ImSim} also derived a modified $m$ that --- in addition to the effects discussed above --- accounted for the impact of blending on the source-galaxy redshift distributions. See their Section 2.3 and 2.4 for more details. \citet[][see their Figure 11]{Amon2022} show that this additional correction has negligible impact on the cosmology constraints given the precision of the DES Y3 sample. Thus, in this work we do not incorporate this additional correction.

Second, our earlier discussions in Section \ref{sec:ImsimPipeline} mentioned that the sky positions of our simulated galaxies do not include cosmological spatial correlations. Section 7 of \citet{Maccrann2022ImSim} discusses in detail the impacts of this approximation for DES Y3 data, utilizing the analysis of \citet{Euclid:2019:Imsims} (who do include such effects) as a reference point. \citet{Maccrann2022ImSim} argue that for DES Y3, this approximation produces a bias whose amplitude is less than the $0.5\%$ estimate from \citet{Euclid:2019:Imsims}. To check for the downstream impact on cosmology, they recommended incorporating an additional 1\% uncertainty on the multiplicative bias priors. \citet[][see their Figure 7]{Amon2022} confirmed cosmology constraints from such an analysis setup are indistinguishable from the fiducial results. Our data is of similar quality to DES Y3, and hence, the same arguments are relevant for our work. We conduct a similar check in Table 4 of \citetalias{paper4}, where we increase our calibration uncertainties by a factor of three (from $\sigma(m) \approx 0.004 \rightarrow 0.012$). The resulting shift in cosmology is within $0.1\sigma$. Thus, the approximation made in our image simulations---of not including spatial clustering in the galaxy positions---has negligible impact on cosmology given the precision of the data. This test also shows that the cosmology results are insensitive to any shifts of $\mathcal{O}(1\%)$ in the multiplicative bias.

\begin{table}[]
    \centering
    \begin{tabular}{c|c|c}
        \hline
          & \textbf{$m \,\,(10^{-2})$ } & \textbf{$c \,\,(10^{-4})$ } \\
        \hline
        Bin 1 & $-0.923 \pm 0.296$ & $-4.966 \pm 1.788$ \\[2pt]
        Bin 2 & $-1.895 \pm 0.421$ & $-5.681 \pm 1.764$\\[2pt]
        Bin 3 & $-4.004 \pm 0.428$ & $-3.812 \pm 1.542$ \\[2pt]
        Bin 4 & $-3.733 \pm 0.462$ & $-7.273 \pm 1.688$\\[2pt]
        Full sample & $-2.454 \pm 0.124$ & $-5.262 \pm 0.959$\\
        \hline
    \end{tabular}
    \caption{The multiplicative calibration factor for the four tomographic bins and the full sample, as determined by the image simulations suite developed in this work. The calibration values are similar to those of DES Y3 \citep{Maccrann2022ImSim}. We also show the additive terms, $c$, which are also similar (but slightly higher than) that work.}
    \label{tab:ShearCalib}
\end{table}

\section{Summary}
\label{sec:summary}

In this paper we present the \decade shape catalog, which contains $107$ million galaxies covering $5,\!412 \deg^2$ in the northern Galactic cap region. The catalog is derived from public DECam imaging data, coming from both large survey programs as well as smaller, standard observing programs. The object shapes are measured using the \textsc{Metacalibration} methodology, and the final source sample is tailored for cosmological analyses with weak lensing. The effective number density of the full source-galaxy sample is $4.56$ galaxies per arcmin$^2$, and the sample has a statistical precision similar to that of the DES Y3 shape catalog \citepalias{y3-shapecatalog}.

To ensure the robustness of this shape catalog, we perform a series of empirical tests including: (1) the mean PSF shape in focal plane coordinates, (2) the residual brighter-fatter effect, (3) the color-dependence of the PSF shape, (4) the mean shear as a function of \textsc{Metacalibration} quantities, (5) the two-point functions of the PSF and galaxy shapes, also known as Rowe statistics and Tau statistics, (6) the tangential shear around stars and field centers, (7) the spatial correlation of the mean shear and various survey properties, and (8) the $B$-mode signals. Overall, we find that the residual systematic effects derived from these tests are either subdominant relative to our signal, or are statistically insignificant. An explicit test of the PSF-related contamination in \citetalias{paper3} also shows the PSF contamination, as estimated from the Rowe and Tau stats, has a completely negligible bias in the derived cosmological constraints. 

In parallel we build an image simulation pipeline to both test our \textsc{Metacalibration} estimator and to quantify the shear calibration bias (and its uncertainty). The latter is done both for the full sample as well as for the four tomographic redshift bins as defined in \citetalias{paper2}. We find an overall shear calibration bias of $m \approx 2.5\%$, with the calibrations for each tomographic bins being within $\approx 1\%$ of this value. These levels of multiplicative biases are comparable to other state-of-the-art weak lensing shear measurements \citep{Li2022, Li2023}, and particularly to the \textsc{Metacalibration}-based measurements of DES Y3 \citep{Maccrann2022ImSim}. The results from our null-tests and the shear calibration indicate it is possible to construct a lensing dataset with similar robustness to that of DES Y3, but while using image data that have significant spatial inhomogeneity and variability in data quality, and that were observed at high airmass. This has implications for the kinds of data that can be incorporated into datasets in upcoming lensing surveys such as Rubin LSST \citep{Abell2009}, the \textit{Nancy Grace Roman Space Telescope} \citep{Spergel:2015:Roman}, and the \textit{Euclid} space telescope \citep{Laureijs:2011:Euclid}.

We briefly note that the coadded images and catalogs created by the \decade cosmic shear project have uses for many astronomical science goals beyond just cosmic shear. In particular, we have assembled a deep, well-tested, multi-band object catalog \citep[see][]{Tan:2024} and value-added products such as a shape catalog (this work) and photometric redshift distributions \citepalias[see][]{paper2}. The coadded images themselves enable a wide suite of science analyses. The \decade shape catalog is publicly available\footnote{\url{dhayaaanbajagane.github.io/data_release/decade}} alongside the coadded images and other object catalogs\footnote{\url{datalab.noirlab.edu/data/delve}} produced within DELVE.

Finally, as mentioned before, this work is part of a series of papers within the \decade project. The others detail the redshift estimates \citepalias{paper2}, the analysis methodology and survey inhomogeneity tests \citepalias{paper3}, and the cosmology results for the \decade catalog \citepalias{paper4}. This work was also limited to image data only in the northern Galactic cap (``DELVE Early Data Release 3''). We have processed public DECam data in the southern Galactic cap, surrounding the DES footprint. These shape measurements (as well as all other necessary calibrations and data products) cover an additional $\approx\!3,\!300 \deg^2$ and are now presented in \href{\#cite.paper5}{Anbajagane \& Chang et al. (\citeyear{paper5})}. The work in this series of papers has set the stage for performing a cosmic shear analysis across $13,\!000 \deg^2$ of the sky, using only DECam data. The full \decade dataset, detailed in \href{\#cite.paper5}{Anbajagane \& Chang et al. (\citeyear{paper5})}, more than doubles the sky coverage of existing, precision weak lensing surveys and therefore, and improves their synergies with many ongoing and upcoming surveys.

\section*{Acknowledgements}

DA is supported by the National Science Foundation (NSF) Graduate Research Fellowship under Grant No.\ DGE 1746045. 
CC is supported by the Henry Luce Foundation and Department of Energy (DOE) grant DE-SC0021949. 
The DECADE project is supported by NSF AST-2108168 and AST-2108169.
The DELVE Survey gratefully acknowledges support from Fermilab LDRD (L2019.011), the NASA {\it Fermi} Guest Investigator Program Cycle 9 (No.\ 91201), and the NSF (AST-2108168, AST-2108169, AST-2307126,  AST-2407526, AST-2407527, AST-2407528). This work was completed in part with resources provided by the University of Chicago’s Research Computing Center. The project that gave rise to these results received the support of a fellowship from "la Caixa" Foundation (ID 100010434). The fellowship code is LCF/BQ/PI23/11970028. C.E.M.-V. is supported by the international Gemini Observatory, a program of NSF NOIRLab, which is managed by the Association of Universities for Research in Astronomy (AURA) under a cooperative agreement with the U.S. National Science Foundation, on behalf of the Gemini partnership of Argentina, Brazil, Canada, Chile, the Republic of Korea, and the United States of America. 

Funding for the DES Projects has been provided by the U.S. Department of Energy, the U.S. National Science Foundation, the Ministry of Science and Education of Spain, 
the Science and Technology Facilities Council of the United Kingdom, the Higher Education Funding Council for England, the National Center for Supercomputing 
Applications at the University of Illinois at Urbana-Champaign, the Kavli Institute of Cosmological Physics at the University of Chicago, 
the Center for Cosmology and Astro-Particle Physics at the Ohio State University,
the Mitchell Institute for Fundamental Physics and Astronomy at Texas A\&M University, Financiadora de Estudos e Projetos, 
Funda{\c c}{\~a}o Carlos Chagas Filho de Amparo {\`a} Pesquisa do Estado do Rio de Janeiro, Conselho Nacional de Desenvolvimento Cient{\'i}fico e Tecnol{\'o}gico and 
the Minist{\'e}rio da Ci{\^e}ncia, Tecnologia e Inova{\c c}{\~a}o, the Deutsche Forschungsgemeinschaft and the Collaborating Institutions in the Dark Energy Survey. 

The Collaborating Institutions are Argonne National Laboratory, the University of California at Santa Cruz, the University of Cambridge, Centro de Investigaciones Energ{\'e}ticas, 
Medioambientales y Tecnol{\'o}gicas-Madrid, the University of Chicago, University College London, the DES-Brazil Consortium, the University of Edinburgh, 
the Eidgen{\"o}ssische Technische Hochschule (ETH) Z{\"u}rich, 
Fermi National Accelerator Laboratory, the University of Illinois at Urbana-Champaign, the Institut de Ci{\`e}ncies de l'Espai (IEEC/CSIC), 
the Institut de F{\'i}sica d'Altes Energies, Lawrence Berkeley National Laboratory, the Ludwig-Maximilians Universit{\"a}t M{\"u}nchen and the associated Excellence Cluster Universe, 
the University of Michigan, NSF's NOIRLab, the University of Nottingham, The Ohio State University, the University of Pennsylvania, the University of Portsmouth, 
SLAC National Accelerator Laboratory, Stanford University, the University of Sussex, Texas A\&M University, and the OzDES Membership Consortium.

The DES data management system is supported by the National Science Foundation under Grant Numbers AST-1138766 and AST-1536171.
The DES participants from Spanish institutions are partially supported by MICINN under grants ESP2017-89838, PGC2018-094773, PGC2018-102021, SEV-2016-0588, SEV-2016-0597, and MDM-2015-0509, some of which include ERDF funds from the European Union. IFAE is partially funded by the CERCA program of the Generalitat de Catalunya.
Research leading to these results has received funding from the European Research
Council under the European Union's Seventh Framework Program (FP7/2007-2013) including ERC grant agreements 240672, 291329, and 306478.
We  acknowledge support from the Brazilian Instituto Nacional de Ci\^encia
e Tecnologia (INCT) do e-Universo (CNPq grant 465376/2014-2).

Based in part on observations at Cerro Tololo Inter-American Observatory at NSF's NOIRLab, which is managed by the Association of Universities for Research in Astronomy (AURA) under a cooperative agreement with the National Science Foundation.

This work has made use of data from the European Space Agency (ESA) mission {\it Gaia} (\url{https://www.cosmos.esa.int/gaia}), processed by the {\it Gaia} Data Processing and Analysis Consortium (DPAC, \url{https://www.cosmos.esa.int/web/gaia/dpac/consortium}).
Funding for the DPAC has been provided by national institutions, in particular the institutions participating in the {\it Gaia} Multilateral Agreement.

This paper is based on data collected at the Subaru Telescope and retrieved from the HSC data archive system, which is operated by the Subaru Telescope and Astronomy Data Center (ADC) at NAOJ. Data analysis was in part carried out with the cooperation of Center for Computational Astrophysics (CfCA), NAOJ. We are honored and grateful for the opportunity of observing the Universe from Maunakea, which has the cultural, historical and natural significance in Hawaii. 

This research uses services or data provided by the Astro Data Lab, which is part of the Community Science and Data Center (CSDC) Program of NSF NOIRLab. NOIRLab is operated by the Association of Universities for Research in Astronomy (AURA), Inc. under a cooperative agreement with the U.S. National Science Foundation.

This manuscript has been authored by Fermi Forward Discovery Group, LLC under Contract No.\ 89243024CSC000002 with the U.S. Department of Energy, Office of Science, Office of High Energy Physics.

All analysis in this work was enabled greatly by the following software: \textsc{Pandas} \citep{Mckinney2011pandas}, \textsc{NumPy} \citep{vanderWalt2011Numpy}, \textsc{SciPy} \citep{Virtanen2020Scipy}, and \textsc{Matplotlib} \citep{Hunter2007Matplotlib}. We have also used
the Astrophysics Data Service (\href{https://ui.adsabs.harvard.edu/}{ADS}) and \href{https://arxiv.org/}{\texttt{arXiv}} preprint repository extensively during this project and the writing of the paper.

\section*{Data Availability}

All catalogs and derived data products (data vectors, redshift distributions, calibrations etc.) for the cosmology analysis are now publicly available through the Noirlab Datalab portal \citep{Fitzpatrick:2014:DataLab, Nikutta:2020:DataLab} as well as through Globus and other avenues. Please visit \url{dhayaaanbajagane.github.io/data_release/decade} for a list of the available dataproducts and their corresponding data access. Our intention is to make all useful products immediately available to the community. Please reach out to DA if a data product of interest to you is not on the above list.

\bibliographystyle{mnras}
\bibliography{References}

\appendix

\section{PSF modeling}
\label{sec:psfex}

\begin{table*}
    \centering
    \begin{tabular}{ll|ll|ll}
        \hline
        PSF model & & PSF variability && Sample selection \\
        \hline
        Parameter & Value & Parameter & Value & Parameter & Value \\
        \hline
        \texttt{BASIS\_TYPE} & PIXEL  & \texttt{PSFVAR\_KEYS}& \texttt{XWIN\_IMAGE,YWIN\_IMAGE} & \texttt{SAMPLE\_AUTOSELECT}& Y\\
        \texttt{BASIS\_NUMBER}& 20 & \texttt{PSFVAR\_GROUPS}& 1,1 & \texttt{SAMPLEVAR\_TYPE}& SEEING\\
        \texttt{BASIS\_NAME} & basis.fits & \texttt{PSFVAR\_DEGREES}& 2 & \texttt{SAMPLE\_FWHMRANGE}& 2.0,15.0\\
        \texttt{BASIS\_SCALE}& 1.0 & \texttt{PSFVAR\_NSNAP}& 9 & \texttt{SAMPLE\_VARIABILITY}& 0.2\\
        \texttt{NEWBASIS\_TYPE} & NONE & \texttt{HIDDENMEF\_TYPE}& COMMON & \texttt{SAMPLE\_MINSN}& 20\\
        \texttt{NEWBASIS\_NUMBER} & 8 & \texttt{STABILITY\_TYPE}& EXPOSURE & \texttt{SAMPLE\_MAXELLIP}& 0.3\\
        \texttt{PSF\_SAMPLING} & 0.7 & &  & \texttt{SAMPLE\_FLAGMASK} & 0x00fe\\
        \texttt{PSF\_PIXELSIZE} & 1.0 &  &  & \texttt{SAMPLE\_IMAFLAGMASK} & 0x00ff\\
        \texttt{PSF\_ACCURACY} & 0.01 & &  & \texttt{BADPIXEL\_FILTER} & N\\
        \texttt{PSF\_SIZE} & 61,61 & &  & \texttt{BADPIXEL\_NMAX} & 0\\
        \texttt{CENTER\_KEYS} & \texttt{XWIN\_IMAGE,YWIN\_IMAGE}&&  & &\\
        \texttt{PSF\_RECENTER} & N &&  & &\\
        \texttt{PHOTFLUX\_KEY} & \texttt{FLUX\_APER(8)}&&  & &\\
        \texttt{PHOTFLUXERR\_KEY} & \texttt{FLUXERR\_APER(8)}&&  & &\\
        \texttt{MEF\_TYPE} & INDEPENDENT &&  & &\\
        \hline
    \end{tabular}
    \caption{Parameter names and values used in the \textsc{PSFEx} run of the DESDM processing pipeline. All shear measurements in this work (and all PSF tests) derive from the PSF model constructed from \textsc{PSFEx}.}
    \label{tab:psfex}
\end{table*}

There are several aspects of the PSF modeling and testing in this work that are different from that of the DES Y3 analysis \citep{Jarvis2020}. We describe here the details of our PSF modeling procedure and our validation that the models are sufficiently accurate for the cosmology analysis.

\subsection{PSF star selection}

The PSF model of a given CCD image is estimated using a subset of stars from that image that are well-suited for constructing the PSF. For example, common considerations are choosing stars that are sufficiently bright to be a good, high signal-to-noise measure of the PSF, but that are not too bright to suffer saturation or other significant non-linearities in the detectors \citep[\eg][]{Jarvis2020}. In this work, the star selection follows the algorithm internal to \textsc{PSFEx}, with the settings shown in Table~\ref{tab:psfex}.

The PSF star catalog is obtained from an initial \textsc{SourceExtractor} run of the CCD image, where this initial run is optimized for finding bright stars and is thus different from the run performed on the coadd images. See \citet[][section 4.5]{Morganson:2018} for more details. This catalog is then postprocessed internally within \textsc{PSFEx}. The first step places a number of cuts on the PSF star properties from \textsc{SourceExtractor}:

\begin{itemize}
    \item \textbf{Signal-to-noise cut:} The algorithm selects \texttt{SNR\_WIN} = \texttt{FLUX\_WIN/FLUXERR\_WIN} $>$ \texttt{SAMPLE\_MINSN}. Here ``\texttt{WIN}'' indicates that the photometry is calculated with a Gaussian window function, where the width of the Gaussian is defined using the object's half-light radius. \vspace{5pt}

    \item \textbf{Size cuts:} \textsc{PSFEx} selects all objects with FWHM in the range given by \texttt{SAMPLE\_FWHMRANGE} (in pixel units). It then iteratively shrinks the upper and lower bounds of this range until the remaining objects have FWHM values that vary within $20\%$. The FWHM used is simply twice the \texttt{FLUX\_RADIUS} as computed by \textsc{SourceExtractor}. \vspace{5pt}

    \item \textbf{Ellipticity cut:} \textsc{PSFEx} removes all objects with ellipticity larger than \texttt{SAMPLE\_MAXELLIP}.  
\end{itemize}

The second selection step occurs during PSF model building. This is performed iteratively, where  
\begin{itemize}
    \item The PSF model is estimated.\vspace{3pt}
    \item The PSF model is compared with the measurements from the PSF stars \vspace{3pt}
    \item If any PSF star exhibits a large $\chi^2$ between its measurement and the model, the star is considered an outlier and removed. \vspace{3pt}
    \item The PSF model is estimated again without these outliers.\vspace{3pt}
    \item The above steps are performed three times in total, with the $\chi^2$ threshold being halved with each iteration. 
\end{itemize}

This selection procedure is somewhat different from DES Y3 and results in a larger PSF star sample that goes to lower signal-to-noise. In DES Y3, the star selection is part of the \textsc{Piff} algorithm \citep{Jarvis2020}. In \textsc{Piff}, after a rough selection in the size and magnitude plane, the algorithm first identifies the brightest 10\% of the objects (excluding saturated objects) and finds a narrow locus at small object-size for the stars and a broad locus of galaxies with larger object sizes. Then the algorithm proceeds to fainter magnitudes, building up both loci, until the stellar locus and the galaxy locus start to merge.

We note that there are two additional factors to consider here. First, in the \textsc{PSFEx} model-fitting step each star is effectively weighted by its signal-to-noise. So, the subsample of faint PSF stars does not contribute as much to the model building, and most of the brighter stars will be those found and used in \textsc{Piff}. Second, due to the unique characteristics of our dataset, there is both (i) more variation in the depth between exposures and (ii) more exposures located closer to the Galactic plane. These two factors contribute to a wider spread in the number of PSF stars in our data as well as a higher number on average. The higher number counts of PSF stars is not an obvious concern for weak lensing analysis, and our PSF tests --- such as the Rowe and Tau statistics (Section~\ref{sec:rowe}) and tangential shear around stars (Section~\ref{sec:gammat_center}) --- show our PSF model is performing adequately.      

\subsection{The PSF model}

\textsc{PSFEx} uses the cutout images of each PSF star selected above in the dimension of \texttt{PSF\_SIZE}, and decomposes them into Principal Component Analysis (PCA) elements -- each star is now represented by a linear combination of Principle Components (PCs), where we keep only a small number (\texttt{NEWBASIS\_NUMBER}) of PCs. Next, \textsc{PSFEx} fits a smooth polynomial function (of order \texttt{PSFVAR\_DEGREES}) across each CCD chip to the coefficients of the PCs. This becomes the PSF model for that chip and the PSF at a given location of a galaxy can be reconstructed given the polynomial fit and the PCs.

\subsection{Reserved stars and PSF tests}\label{appx:PSFStarSample}

In DES Y3, once the star selection is done, a random 20\% subset of that selection is set aside and not used for the PSF model. This subset serves as a clean sample to test the quality of the PSF model since the objects were not used in constructing the PSF model. In our work, we do not do this step as \textsc{PSFEx} does not provide such a reserve star sample. Furthermore, it was found in DES that the PSF tests (of the kind we perform) are insensitive when swapping the reserved stars sample with the full sample. This is likely because the limiting factor for the quality of the PSF model is usually not the number of stars, but rather the flexibility of the model itself.

To reduce the noise in statistics derived from the $\boldsymbol{e}^{\star}$ and $ T^{\star}$ measurements (\eg the Rowe and Tau statistics in Section~\ref{sec:rowe}), we place a signal-to-noise cut on the PSF star sample at $\texttt{SNR\_WIN} > 80$. Figure~\ref{fig:Rowe_stats} shows an example of variation in the signal-to-noise cut and shows it does not change the qualitative findings of our PSF tests. In addition, our PSF tests (all shown in Section~\ref{sec:sys_tests}) use a subsampled PSF star catalog rather than the entire catalog. The full catalog is $N \approx 10^9$ in size, and this significantly increases the computational expense of our tests. The PSF tests can be accurately performed with far fewer stars; for example, the DES Y3 tests used $N \sim 10^7$ stars for their tests. In our tests, we reduce the sample size by randomly subsampling 20\% of the objects. This subsample is used for all tests and is referred to as the PSF star sample in Section ~\ref{sec:sys_tests}.



\appendix


\label{lastpage}
\end{document}